\documentclass[12pt,letterpaper]{article}

\usepackage{amsmath,amssymb,calc}
\usepackage{graphicx}

\newcommand\be{\begin{equation}}
\newcommand\ee{\end{equation}}
\newcommand{\bea}{\begin{eqnarray}}
\newcommand{\eea}{\end{eqnarray}}

\newcommand{\nn}{\nonumber}
\newcommand{\pd}{\partial}

\newcommand{\la}{\langle}
\newcommand{\ra}{\rangle}
\def\id{\protect{{1 \kern-.28em {\rm l}}}}

\font\mybb=msbm10 at 12pt

\def\bb#1{\hbox{\mybb#1}}

\def\id{\protect{{1 \kern-.28em {\rm l}}}}
\def\unit{\relax{\rm 1\kern-.26em I}}

\begin{document}

\begin{titlepage}
\begin{center}
\hfill ITP-UU-08/36 \\
\hfill SPIN-08/27 \\
\vskip 15mm

{\Large {\bf Flux Vacua Attractors and Generalized Compactifications \\[3mm] }}

\vskip 10mm

{\bf Lilia Anguelova}

\vskip 4mm
{\em Institute for Theoretical Physics and Spinoza Institute}\\
{\em Utrecht University, 3508 TD Utrecht, The Netherlands}\\
{\tt L.Anguelova@phys.uu.nl}\\

\vskip 6mm

\end{center}

\vskip .1in
\vspace{2cm}

\begin{center} {\bf ABSTRACT }\end{center}

\begin{quotation}\noindent

We investigate whether there are attractor equations for $N=1$ flux vacua in generalized compactifications. We fill a gap in the existing literature by verifying analytically that the recently proposed susy attractors, for type IIB CY(3) orientifold compactifications with flux, do give supersymmetric minima of the relevant scalar potential. Furthermore, our considerations clarify various confusions about existing proposals for generalization of the flux vacua attractors to non-K\"{a}hler compactifications. We explore different possibilities for generalization and find attractor equations for $N=1$ {\it Minkowski vacua only} both for the heterotic string on $SU(3)$ structure manifolds and for type IIA/B on $SU(3)\times SU(3)$ structure spaces.

\end{quotation}
\vfill

\end{titlepage}

\eject

\tableofcontents

\section{Introduction}

The attractor mechanism was first discovered in the context of supersymmetric black holes (BH) in $N=2$, $d=4$ supergravity coupled to vector multiplets \cite{FKS, FK, AS}. In essence, it means that the near horizon geometry of the black holes does not change under smooth variations of the asymptotic values of the various scalar fields (moduli). In particular, the values of the moduli at the horizon are independent of the asymptotic values of the scalars and, instead, are completely determined by the BH electric and magnetic charges via a set of algebraic equations, called attractor equations. Subsequently, this kind of behaviour was established for black holes in various extended supergravities and in different dimensions. For recent reviews see \cite{FHM, BFKM, FL}; the enormous amount of literature on the topic is referenced therein. 

It was realized in \cite{RK} that there is a technical similarity between the situation for BH attractors in $N=2$, $d=4$ and flux vacua compactifications of type IIB on CY(3) with O(3)/O(7) planes. More precisely, the relevant scalar potentials in the two cases have very similar forms and, in addition, in both cases the moduli spaces have special K\"{a}hler geometry; although the latter property is not present for a generic $N=1$, $d=4$ supergravity, in this context it is inherited from the $N=2$ theory, obtained from IIB on a CY 3-fold without orientifolding. Inspired by this similarity, Kallosh argued that in the IIB case one can also write down attractor equations or, in other words, that for IIB CY(3) orientifolds the minimization of the effective $N=1$ supergravity potential is equivalent to solving a simplified system of attractor equations.

Apart from being an interesting observation, the flux vacua attractors of \cite{RK} are potentially of significant importance for the problem of moduli stabilization in string theory. The latter is a long-standing problem whose resolution has only started taking shape in recent years. The issue is that string compactifications on CY 3-folds have many allowed deformations of the internal space, which do not cost any energy. Those deformations manifest themselves in 4d as scalar fields without a potential, called moduli. Since the various parameters of the four-dimensional effective theory (coupling constants, for example) depend on those moduli, the arbitrary values that the latter can take lead to lack of predictability of string theory. 

This problem can be resolved by considering background fluxes \cite{DRS, GKP} and perturbative  \cite{Quevedo} and/or non-perturbative effects \cite{KKLT} that induce a potential for the moduli. For a comprehensive review on flux compactifications, see \cite{Grana}. However, generically turning on background fluxes deforms the internal manifold away from Calabi-Yau. The resulting internal geometry can be described most efficiently in the language of $SU(3)\times SU(3)$ structures. These are the most general type II compactifications that give an effective $N=2$ theory in 4d \cite{GLW1, GLW2}. The $N=1$ vacua in this context, studied in \cite{GMPT}, are a vast generalization of the IIB orientifolds considered in \cite{RK}. Unlike the IIB orientifolds though, the superpotential in these generalized compactifications depends on all relevant geometric moduli and so, in principle, it is possible to achieve moduli stabilization at the classical level. 

$SU(3)\times SU(3)$ structure spaces are characterized by a pair of pure $SO(6,6)$ spinors $\Phi_+$ and $\Phi_-$, which for a CY(3) reduce to the familiar K\"{a}hler form $J$ and holomorphic 3-form $\Omega$. In general $\Phi_+$ ($\Phi_-$) is a sum of even (odd) forms of different degrees. It was shown in \cite{NH, GLW1, GLW2, CB} that the deformation spaces of $\Phi_+$ and $\Phi_-$ have special K\"{a}hler geometry. So it is a natural question to ask whether the new attractors of \cite{RK} can be extended to (at least some cases) of type II on $SU(3)\times SU(3)$ structure spaces. A positive answer would be of great value for the study of moduli stabilization in these compactifications. The reason is that generically the attractor equations are (significantly) simpler than the conditions one obtains directly from minimizing the scalar potential. In this paper we will explore whether there are attractors for $N=1$ flux vacua in those generalized type II compactifications.

A more familiar special case of generalized geometry is provided by spaces with $SU(3)$ structure. Type II on such non-K\"{a}hler manifolds can be obtained from the $SU(3)\times SU(3)$ structure case by taking the diagonal $SU(3)$ subgroup. For the heterotic string, however, the $SU(3)$ structure compactifications are the most general ones, since the heterotic supersymmetry transformations are only sensitive to a single internal spinor.\footnote{I.e., even if the internal manifold has $SU(3)\times SU(3)$ structure, the heterotic string still feels only a single linear combination of the two internal spinors and so is effectively compactified only on an $SU(3)$ structure manifold. We thank D. Waldram for a valuable discussion on this point.} Clearly, heterotic non-Kahler compactifications provide another natural context, in which to search for generalizations of the flux vacua attractors of \cite{RK}. In fact, this issue was already addressed in \cite{GD}. Although that reference has a very good intuition, we disagree with the details. The reason things are more subtle than anticipated in \cite{GD}, is the following. A key ingredient in Kallosh's argument for IIB attractors is the generic expansion, derived in \cite{DD}, of a suitable four-form flux in terms of basic geometric structures of a related CY(4). Now, generalized geometry structures are in principle much less understood than CY manifolds. And, in particular, the analogue of the expansion of \cite{DD} for the generalized case has not been established yet. So the approach to follow is to make a proposal and then verify that it works. In order to accomplish the latter step, it is very helpful to understand analytically how the new attractors satisfy the relevant supersymmetry conditions.\footnote{Throughout this paper we only consider supersymmetric attractors.} So far, this has only been checked by numerical methods \cite{AG} and, as a result, it was not particularly clear which properties of the IIB CY orientifolds are essential and which are accidental regarding the existence of the new attractors. We perform an explicit analytical verification that the flux vacua attractors of \cite{RK} give supersymmetric minima of the relevant scalar potential. In doing so, we are led to the realization that a necessary condition is not satisfied in the single-K\"{a}hler-modulus conjecture of \cite{GD} for heterotic non-K\"{a}hler attractors. Despite that, we show that a slight modification of another proposal of \cite{GD} gives heterotic attractors for Minkowski vacua in the case of arbitrary number of K\"{a}hler moduli.\footnote{Note that at the classical level the heterotic string has only Minkowski flux vacua, unlike type II strings. In order to obtain heterotic AdS vacua, one has to include quantum effects like, for example, gaugino condensation; see \cite{FLi}.} Furthermore, we also find flux vacua attractor equations for type IIA/B on $SU(3)\times SU(3)$ structure, although {\it only} for Minkowski vacua. 

The present paper is organized as follows. In Section 2 we give some useful background material. In Subsection 2.1 we collect a few necessary properties of special K\"{a}hler geometry and in Subsection 2.2 we briefly review the well-known BH attractors. In Section 3 we explain in more detail the motivation for and the derivation of the flux vacua attractors in type IIB CY(3) orientifold compactifications \cite{RK}. In Section 4 we verify analytically that these attractor equations give solutions of the relevant supersymmetry conditions. In Section 5 we explore the possibilities for generalization. In Subsection 5.1 we study the heterotic string on non-K\"{a}hler manifolds. After reviewing some necessary properties of these compactifications, we scrutinize the conjecture of \cite{GD} and we establish the existence of heterotic attractors for Minkowski vacua. In Subsection 5.2 we investigate type IIA/B on $SU(3)\times SU(3)$ structure spaces. We start by reviewing some necessary properties. Then we recall the known results for the superpotential and K\"{a}hler potential of generic $N=1$ trunctions and find supersymmetric attractor equations, for Minkowski vacua only. Finally, in Appendix A we present a technical computation, which completes the considerations of Section 4.

\section{Preliminaries} \label{SKG}
\setcounter{equation}{0}

In this section we review some properties of special K\"{a}hler geometry, that will be needed throughout the paper. Also, as a useful preparation for the flux vacua attractors, we outline the derivation of the attractor equations for black holes in $N=2$, $d=4$ supergravity coupled to $n$ vector multiplets.

\subsection{Special K\"{a}hler geometry}

In a coordinate-independent way, a $2n$-real-dimensional special K\"{a}hler manifold is defined in terms of a symplectic section
\be
(L^{\Lambda}, M_{\Lambda}) \, , \qquad \Lambda =0,1,...,n
\ee
of an $Sp(2n+2)$ vector bundle over a Hodge-K\"{a}hler manifold, such that
\be \label{Norm}
i (\bar{L}^{\Lambda} M_{\Lambda} - L^{\Lambda} \bar{M}_{\Lambda}) = 1 \, .
\ee
$L^{\Lambda}$ and $M_{\Lambda}$ are covariantly holomorphic functions of the coordinates $\{t^i, \bar{t}^{\bar{i}} \}$, also called moduli, that parametrize the special K\"{a}hler manifold. In other words, 
\be
D_{\bar{i}} L^{\Lambda} = (\pd_{\bar{i}} - \frac{1}{2} K_{\bar{i}}) L^{\Lambda} = 0 \, ,
\ee
where clearly $D_{\bar {i}}$ is the K\"{a}hler covariant derivative. Also, $D_i L^{\Lambda} = (\pd_i + \frac{1}{2} K_i) L^{\Lambda}$ and similarly for $M_{\Lambda}$. The relation between $(L^{\Lambda} (t), M_{\Lambda} (t))$ and the holomorphic section $(X^{\Lambda} (t), G_{\Lambda} (t))$ is the following:
\be \label{LXMG}
L^{\Lambda} = e^{K/2} X^{\Lambda} \, , \qquad M_{\Lambda} = e^{K/2} G_{\Lambda} \, ,
\ee
where $\pd_{\bar{i}} X^{\Lambda} = 0$ and $\pd_{\bar{i}} G_{\Lambda} = 0$. The K\"{a}hler potential is given by 
\be
K = - \ln i (\bar{X}^{\Lambda} G_{\Lambda} - X^{\Lambda} \bar{G}_{\Lambda}) \, ,
\ee
as can be seen from (\ref{Norm}) and (\ref{LXMG}). From special geometry one has:
\be \label{MNL}
M_{\Lambda} = {\cal N}_{\Lambda \Sigma} L^{\Sigma} \, , \qquad D_i M_{\Lambda} = \overline{{\cal N}}_{\Lambda \Sigma} D_i L^{\Sigma} \, ,
\ee
where ${\cal N}_{\Lambda \Sigma}$ is a complex symmetric $(n+1)\times (n+1)$ matrix. Note that the K\"{a}hler weight of ${\cal N}_{\Lambda \Sigma}$ is $0$, meaning that $D_i {\cal N}_{\Lambda \Sigma} = \pd_i {\cal N}_{\Lambda \Sigma}$. We will also need the following relations \cite{CDFvP}:
\be \label{NG}
{\cal N}_{\Lambda \Sigma} = \bar{G}_{\Lambda \Sigma} + 2 i \frac{({\rm Im} G_{\Lambda \Gamma}) ({\rm Im} G_{\Sigma \Pi}) L^{\Gamma} L^{\Pi}}{({\rm Im} G_{\Xi \Omega}) L^{\Xi} L^{\Omega}}
\ee
and
\be \label{ImGLDL}
{\rm Im} G_{\Lambda \Sigma} \,L^{\Lambda} \,D_{\bar{i}} \bar{L}^{\Sigma} = 0 \, .
\ee
For more details on special K\"{a}hler geometry see \cite{CDFvP}.

\subsection{BH attractors} \label{BHat}

Let us now briefly recall the well-known BH attractor equations \cite{FK}. To derive them, one considers the central charge
\be \label{BHCC}
Z = q_{\Lambda} L^{\Lambda} - p^{\Lambda} M_{\Lambda} \, ,
\ee
which is a function of the moduli $t^i,\bar{t}^{\bar{i}}$ and the black hole electric and magnetic charges $p^{\Lambda}$ and $q_{\Lambda}$ respectively. This function determines the black hole potential via 
\be \label{VBH}
V_{BH} = |Z|^2 + |D_i Z|^2 \, .
\ee
The supersymmetric critical points of $V_{BH}$ are given by the solutions of 
\be
D_i Z = \left( \pd_i + \frac{1}{2} K_i \right) Z = 0 \, .
\ee
Using that $\pd_{\bar{i}} {\cal N}_{\Lambda \Sigma} \,L^{\Sigma} = 0$ \cite{CDFvP}, one can solve the central charge minimization condition
\be
D_{\bar{i}} \bar{Z} = q_{\Lambda} D_{\bar{i}} \bar{L}^{\Lambda} - p^{\Lambda} {\cal N}_{\Lambda \Sigma} D_{\bar{i}} \bar{L}^{\Sigma} = 0
\ee
for the charges in terms of the moduli. The result is the BH attractor equations \cite{FK}:
\be \label{AttrEqBH}
p^{\Lambda} = i (\bar{Z} L^{\Lambda} - Z \bar{L}^{\Lambda}) \, , \qquad q_{\Lambda} = i (\bar{Z} M_{\Lambda} - Z \bar{M}_{\Lambda}) \, ,
\ee
where the right hand side is understood to be evaluated at the BH horizon (more precisely, at a fixed point in the near horizon geometry). Clearly, this also implies that the values of the moduli at the horizon are fixed in terms of the BH charges.

\section{New attractors}
\setcounter{equation}{0}

In \cite{RK}, Kallosh formulated a generalization of the attractor equations (\ref{AttrEqBH}) for the case of type IIB CY(3) orientifold compactifications. The motivation for this is the following. The supergravity potential for the effective $N=1$, $d=4$ description, 
\be \label{VFC}
V = |DZ|^2 - 3 |Z|^2 \qquad {\rm with} \qquad Z = e^{\frac{K}{2}} W \, ,
\ee
is very similar to (\ref{VBH}) and, furthermore, the $N=1$ theory inherits certain special geometry properties from the $N=2$ one, obtained by compactifying type IIB on a CY(3). In order to be able to state the new attractor equations, let us first recall a few useful facts about the special K\"{a}hler geometry of type IIB CY compactifications.

\subsection{Type IIB on a CY(3)}

We denote by $\alpha_a$ and $\beta^a$ the basis of CY 3-forms and by $(A^a,B_a)$ their dual 3-cycles. Then the holomorphic sections are given by
\be
X^a (z) = \int_{A^a} \Omega \qquad {\rm and} \qquad G_a (z) = \int_{B_a} \Omega \, ,
\ee
where $\Omega$ is the holomorphic 3-form of the CY manifold and $z^i$ are the complex structure moduli.\footnote{As in \cite{RK}, we use different notation for the indices and moduli in the IIB CY orientifold case, in order to distinguish it from the general discussion in Section \ref{SKG}.} In terms of an expansion in $(\alpha_a, \beta^a)$, one can write:
\be \label{Om}
\Omega = X^a \alpha_a - G_a \beta^a \, .
\ee
The K\"{a}hler potential for the complex structure moduli of the CY is
\be \label{KXG}
K = - \ln \,i \int \Omega \wedge \bar{\Omega} = - \ln i \left( \bar{X}^a G_a - X^a \bar{G}_a \right) .
\ee

Before turning to the flux vacua attractor equations of \cite{RK}, it is useful to recall a derivation of the BH attractors in the context of black holes in IIB compactifications on CY(3) \cite{AS, RK}.\footnote{This is different from the original derivation in \cite{FK}.} In this context the central charge is
\be
Z = e^{\frac{K}{2}} \int H \wedge \Omega \, ,
\ee
where $H$ is the NS 3-form flux. We can expand the latter in the $(\alpha_a, \beta^a)$ basis:
\be \label{Hexp}
H = p^a \alpha_a - q_a \beta^a \, .
\ee
However, we can also write $H$ in the following way:
\be \label{HD}
H = i [\bar{Z} \hat{\Omega} - K^{i\bar{j}} (D_{\bar{j}} \bar{Z}) D_i \hat{\Omega}] + c.c. \,\, ,
\ee
where $\hat{\Omega} = e^{\frac{K}{2}} \Omega$ and there are no terms proportional to $D_i D_j \hat{\Omega}$ because the latter can be expressed in terms of $\bar{D}_{\bar{i}} \bar{\hat{\Omega}}$ due to special geometry. Now, since in a supersymmetric minimum $D_i Z = 0$, for black hole attractors the above expression for $H$ becomes:
\be \label{Hattr}
H = i ( \bar{Z} \hat{\Omega} - Z \bar{\hat{\Omega}} ) \, .
\ee
Using (\ref{Om}), (\ref{LXMG}) and (\ref{Hexp}), one can immediately see that (\ref{Hattr}) gives precisely the attractor equations (\ref{AttrEqBH}).

\subsection{Flux vacua attractors}

The idea in \cite{RK} is to use for flux compactifications an expansion of the fluxes analogous to (\ref{HD}) and, by evaluating its right-hand side at the flux vacua, to obtain attractor equations like (\ref{Hattr}). The equations, that one obtains in this way, are then equivalent to the minimization of the scalar potential in this class of flux compactifications. More precisely, \cite{RK} considers type IIB on a CY(3) with O3/O7 planes. In this case, the RR and NS 3-form fluxes can be decomposed as:
\be
H = p_h^a \alpha_a - q_{ha} \beta^a \qquad {\rm and} \qquad F = p_f^a \alpha_a - q_{fa} \beta^a \, .
\ee 
The superpotential of the effective $N=1$ four-dimensional theory is:
\be \label{Sup}
W = \int (F - \tau H) \wedge \Omega = (q_{fa} - \tau q_{ha}) X^a - (p_f^a - \tau p_h^a) G_a \, ,
\ee
where $\tau$ is the axion-dilaton. The K\"{a}hler potential for $\tau$ and for the complex structure moduli is:
\be \label{Ktot}
K = - \ln [-i (\tau - \bar{\tau})] - \ln [i \int \Omega \wedge \bar{\Omega}] \, .
\ee 
Similarly to before, one takes the central charge to be:
\be
Z = e^{\frac{K}{2}} W = e^{\frac{K}{2}} \int [F\wedge \Omega + H \wedge (-\tau \Omega )] \, .
\ee

In order to write attractor equations, one needs to extend special geometry to the moduli space containing both the axion-dilaton and the complex structure moduli. To do that, Kallosh introduces new symplectic sections,
\be
\Xi = \left( \begin{array}{c}
\Xi_1 \\
\Xi_2 \\
\end{array}
\right) =
\left( \begin{array}{c}
\Omega \\
-\tau \Omega \\
\end{array}
\right),
\ee
in terms of which one can write:
\be \label{KXi}
K = - \ln \left[ \int ( \Xi_1 \wedge \bar{\Xi}_2 - \Xi_2 \wedge \bar{\Xi}_1 ) \right] \qquad {\rm and} \qquad Z = e^{\frac{K}{2}} \int (F \wedge \Xi_1 + H \wedge \Xi_2) \, .
\ee
Now, using that at a supersymmetric minimum $D_A Z = 0$ with $A=\tau, i$ and taking $D_A D_B Z = 0$ as in special geometry, Kallosh infers that the attractor equations for flux vacua, analogous to (\ref{Hattr}), are:
\be \label{AE1}
\left( 
\begin{array}{c}
p_h^a \\
q_{ha} \\
p_f^a \\
q_{fa} \\
\end{array}
\right) = e^K
\left(
\begin{array}{c}
\bar{W} X^a + W \bar{X}^a \\
\bar{W} G_a + W \bar{G}_a \\
\tau \bar{W} X^a + \bar{\tau} W \bar{X}^a \\
\tau \bar{W} G_a + \bar{\tau} W \bar{G}_a \\
\end{array}
\right),
\ee
where the right-hand side is understood to be evaluated at the susy minima. However, as pointed out in \cite{RK}, the extended moduli space containing the axion-dilaton is described by special geometry only partially. More precisely, $D_A D_B Z$ does not have to vanish in general. When $D_A D_B Z \neq 0$, it is useful to introduce the following notation:
\be 
F_4 = -\alpha \wedge F + \beta \wedge H \, , \qquad \int_{T^2} \alpha \wedge \beta = 1 \, ,
\ee
where $T^2$ is the auxiliary torus in the F-theory description of type IIB vacua (i.e., as F-theory compactifications on CY 4-folds of the form $CY(4) = (CY(3)\times T^2)/ \bb{Z}_2$\,). Then the expansion
\be \label{F4exp}
F_4 = \bar{Z} \hat{\Omega}_4 - \bar{D}^A \bar{Z} D_A \hat{\Omega}_4 + \bar{D}^{\underline{0}I} \bar{Z} D_{\underline{0}I} \hat{\Omega}_4 + c.c. \,\, ,
\ee
derived in \cite{DD} (see also \cite{MS}), implies according to \cite{RK} the generalized attractor equations:
\be \label{AE2}
\left( 
\begin{array}{c}
p_h^a \\
q_{ha} \\
p_f^a \\
q_{fa} \\
\end{array}
\right) = 
\left(
\begin{array}{c}
\bar{Z} L^a + Z \bar{L}^a \\
\bar{Z} M_a + Z \bar{M}_a \\
\tau \bar{Z} L^a + \bar{\tau} Z \bar{L}^a \\
\tau \bar{Z} M_a + \bar{\tau} Z \bar{M}_a \\
\end{array}
\right) +
\left(
\begin{array}{c}
\bar{Z}^{\underline{0} I} D_I L^a + Z^{\underline{0} I} \bar{D}_I \bar{L}^a \\
\bar{Z}^{\underline{0}I} D_I M_a + Z^{\underline{0}I} \bar{D}_I \bar{M}_a \\
\bar{\tau} \bar{Z}^{\underline{0} I} D_I L^a + \tau Z^{\underline{0} I} \bar{D}_I \bar{L}^a \\
\bar{\tau} \bar{Z}^{\underline{0} I} D_I M_a + \tau Z^{\underline{0} I} \bar{D}_I \bar{M}_a \\
\end{array}
\right).
\ee
As before, the right-hand side is understood to be evaluated at the supersymmetric flux vacua. Also, here $A= (\underline{0},I)$ are flat indices associated with the orthonormal frame $e^{\hat{a}}_A$ such that $ g_{\hat{a}\bar{\hat{b}}} e^{\hat{a}}_A e^{\bar{\hat{b}}}_{\bar{B}} = \delta_{A \bar{B}}$, where the curved indices are $\hat{a}=(0,i)$ with $0$ corresponding to $\tau$ and $i$ to the complex structure moduli. In particular, $Z^{\underline{0} I} = D^{\underline{0}} D^I Z$ with $D^{\underline{0}} = e_{\tau}^{\underline{0}} D^{\tau}$ and $D^I = e_i^I D^i$. In (\ref{F4exp}), $\hat{\Omega}_4$ is given by $\hat{\Omega}_4 = \hat{\Omega}_1 \wedge \hat{\Omega}$, where $\hat{\Omega} = e^{\frac{K_{CS}}{2}} \Omega$ is the covariantly holomorphic 3-form of the CY(3) and $\hat{\Omega}_1$ is a covariantly holomorphic 1-form on the $T^2$. Finally, in (\ref{AE2}) $L^a = e^{K/2} X^a$ with $K$ given in (\ref{Ktot}) and similarly for $M_a$. Thus, unlike (\ref{Norm}), now we have $i (\bar{L}^a M_a - L^a \bar{M}_a) = e^{K_{(\tau)}}$, where $K_{(\tau)} = -\ln [-i(\tau - \bar{\tau})]$. For a more detailed review of the new attractors of \cite{RK}, see \cite{BFKM}.

As already mentioned in the introduction, it will be of great use for the generalization to heterotic on $SU(3)$ structure and type II on $SU(3)\times SU(3)$ structure spaces to have an analytical understanding of how the new attractor equations (\ref{AE1}) and (\ref{AE2}) give solutions to the supersymmetry conditions $DZ = 0$. To the best of our knowledge, such an explicit analytical verification has not been written down so far in the literature (for a numerical check see \cite{AG}). In the next section we will fill this gap.

\section{Verifying new attractors} \label{AttrProof}
\setcounter{equation}{0}

In this section we will show by analytical means that the new attractor equations imply automatically the supersymmetry conditions $D_A Z = 0$, where $A = z^i, \tau$. Although this is just a confirmation of the derivation of \cite{RK}, it will be of great use for the generalizations that we will turn to in Section \ref{Gen}. Now, in the $D_A D_B Z = 0$ case we will see that the flux vacua attractors can be derived in pretty much the same way as the BH attractor equations. However, the general $D_A D_B Z \neq 0$ case is more involved and we will only be able to verify that, when the new attractors hold, the supersymmetry conditions are satisfied. In both cases, though, it will become clear that, in addition to special K\"{a}hler geometry, there is another property that is of crucial importance. Namely, this is the particular form of the K\"{a}hler potential for the axion-dilaton.

\subsection{$D_A D_B Z = 0$ case} \label{NewAttrSimple}

Let us first consider (\ref{AE1}). To see how these equations follow from (\ref{KXi}), let us rewrite the K\"{a}hler potential in (\ref{KXi}) in such a way that it acquires the same form as in (\ref{KXG}):
\bea \label{Ktilde}
K &=& - \ln [\bar{X}^a (\tau G_a) + X^a (\bar{\tau} \bar{G}_a) - (\tau X^a) \bar{G}_a - (\bar{\tau} \bar{X}^a) G_a] \nn \\
&=& - \ln [i (\bar{X}^a \tilde{G}_a - X^a \bar{\tilde{G}}_a) + i (\bar{\tilde{X}}^a G_a - \tilde{X}^a \bar{G}_a)] \, ,
\eea
where we have introduced the notation $\tilde{X}^a = -i \tau X^a$ and $\tilde{G}_a = -i \tau G_a$. Note that, unlike the BH case in which $X^a$ and $G_a$ were paired, now the pairs of sections are $(X^a, \tilde{G}_a)$ and $(\tilde{X}^a, G_a)$. In terms of these, the superpotential (\ref{Sup}) becomes:
\be
W = q_{fa} X^a - (-i p_h^a) \tilde{G}_a + (-i q_{ha}) \tilde{X}^a - p^a_f G_a \, .
\ee
It is clear now that for the doubled sections $X^{a'} = (X^a, \tilde{X}^a)$ and $G_{a'} = (\tilde{G}_a, G_a)$ and the electric and magnetic charge vectors $q_{a'} = (q_{fa}, -i q_{ha})$ and $p^{\,a'} = (-i p_h^a, p^a_f)$ one has exactly the same situation as for the original BH attractors, Subection \ref{BHat}. Hence, one can derive the generalized attractor equations (\ref{AE1}) from the susy conditions $D_i Z = 0$ in exactly the same way that equations (\ref{AttrEqBH}) were derived in \cite{FK}. One finds:\footnote{Clearly, the derivation of \cite{FK} is modified (in an obvious way) for the purely imaginary charges $-i p_h^a$ and $-i q_{ha}$.}
\be \label{AT}
\left( 
\begin{array}{c}
-i p_h^a \\
q_{fa} \\
p_f^a \\
-i q_{ha} \\
\end{array}
\right) = i \,e^K
\left(
\begin{array}{c}
-(\bar{W} X^a + W \bar{X}^a) \\
\,\, \bar{W} \tilde{G}_a - W \bar{\tilde{G}}_a \\
\,\, \bar{W} \tilde{X}^a - W \bar{\tilde{X}}^a \\
-(\bar{W} G_a + W \bar{G}_a) \\
\end{array}
\right).
\ee
Substituting $\tilde{X}^a = -i \tau X^a$, $\tilde{G}_a = -i \tau G_a$ in the above expressions and rearranging the order of the rows, one obtains exactly (\ref{AE1}).

This derivation is so simple that one may be tempted to think that the generalization of the BH attractor equations to $N=1$ flux vacua in type IIB is a mere triviality. However, once we turn to the more general case, namely equations (\ref{AE2}), it will be clear that this is not so. In fact, one can already see at this stage that things are nontrivial by noticing that we have obtained (\ref{AE1}) without using the $D_{\tau} Z = 0$ equation. So we still need to show that the latter is satisfied when equations (\ref{AE1}) hold. To do that, let us first write the central charge as:
\bea \label{Znew}
Z = e^{\frac{K}{2}} \int (F-\tau H) \wedge \Omega &=& (q_{fa} - \tau q_{ha}) L^a - (p_f^a - \tau p_h^a) M_a \nn \\
&=& q_{fa} L^a - i q_{ha} \tilde{L}^a -p_f^a M_a + ip_h^a \tilde{M}_a \, ,
\eea
where we have defined $\tilde{L}$ and $\tilde{M}$ in the same way as $\tilde{X}$ and $\tilde{G}$ above, i.e. $\tilde{L}^a = -i\tau L^a$ and $\tilde{M}_a = -i \tau M_a$. Then we have:
\be \label{DtauZ}
D_{\tau} Z = q_{fa} D_{\tau} L^a - i q_{ha} D_{\tau} \tilde{L}^a -p_f^a D_{\tau} M_a + ip_h^a D_{\tau} \tilde{M}_a \, .
\ee 
Note that $L^a (\tau, z^i)$ and $M_a (\tau, z^i)$ are related to $X^a (z^i)$ and $G_a (z^i)$, respectively, via the total K\"{a}hler potential (\ref{Ktot}). Therefore 
\be
D_{\tau} L^a = \left( \pd_{\tau} + \frac{1}{2} K_{\tau} \right) L^a = K_{\tau} L^a
\ee
and similarly $D_{\tau} M_a = K_{\tau} M_a$. Using the latter relations, together with $D_{\tau} \tilde{L}^a = -i L^a - i \tau D_{\tau} L^a$ and $D_{\tau} \tilde{M}_a = -i M_a -i \tau D_{\tau} M_a$, and substituting the charges from (\ref{AE1}) we find:
\bea \label{Dtau}
D_{\tau} Z &=& (\tau \bar{Z} M_a + \bar{\tau} Z \bar{M}_a) K_{\tau} L^a - (\bar{Z} M_a + Z \bar{M}_a) (1 + \tau K_{\tau}) L^a \nn \\
&-& (\tau \bar{Z} L^a + \bar{\tau} Z \bar{L}^a) K_{\tau} M_a + (\bar{Z} L^a + Z \bar{L}^a) (1+ \tau K_{\tau}) M_a \nn \\
&=& - i Z (1 - \bar{\tau} K_{\tau} + \tau K_{\tau}) e^{K_{(\tau)}} \, ,
\eea
where in the second equality we have used that $i(\bar{L}^a M_a - L^a \bar{M}_a) = e^{K_{(\tau)}}$ with $K_{(\tau)}$ being the $\tau$-dependent part of the K\"{a}hler potential (\ref{Ktot}), i.e. $K_{(\tau)} = -\ln [-i (\tau - \bar{\tau})]$. Hence, making use of
\be \label{Ktau}
K_{\tau} = - \frac{1}{\tau - \bar{\tau}} \, ,
\ee
we find that (\ref{Dtau}) gives $D_{\tau} Z = 0$, which is exactly what we wanted to show.

Note that relation (\ref{Ktau}) is essential in the proof that the new attractor equations give solutions to the susy condition $D_{\tau} Z = 0$. In other words, the special geometry inherited by the $N=1$ theory is not enough by itself, contrary to previous expectations in the literature. This observation is of crucial importance regarding proposed generalizations of the flux vacua attractors of \cite{RK} to heterotic non-K\"{a}hler compactifications \cite{GD}. We will elaborate further on this in Section \ref{Gen}.

\subsection{$D_A D_B Z \neq 0$ case} \label{ProofGen}

Now let us turn to the generalized equations (\ref{AE2}). It is not clear to us how to extend  the argument presented in (\ref{Ktilde})-(\ref{AT}) to the case $D_A D_B Z \neq 0$. So, instead of trying to obtain (\ref{AE2}) from $DZ=0$, we will simply verify that, when equations (\ref{AE2}) hold, the susy minimum conditions $D_i Z = 0$ and $D_{\tau} Z=0$ are automatically satisfied. In that regard, it is instructive to see first how that works for the original equations (\ref{AttrEqBH}).

For easier comparison with \cite{FK}, let us consider $D_{\bar{i}} \bar{Z}$ instead of $D_i Z$. Substituting the charges from (\ref{AttrEqBH}), we find:
\be \label{DZ}
D_{\bar{i}} \bar{Z} = q_{\Lambda} D_{\bar{i}} \bar{L}^{\Lambda} - p^{\Lambda} {\cal N}_{\Lambda \Sigma} D_{\bar{i}} \bar{L}^{\Sigma} = -i Z (\bar{M}_{\Lambda} - \bar{L}^{\Sigma} {\cal N}_{\Sigma \Lambda}) D_{\bar{i}} \bar{L}^{\Lambda} \, .
\ee
Since ${\cal N}_{\Lambda \Sigma}$ is complex, in other words ${\cal N}_{\Lambda \Sigma} \bar{L}^{\Sigma} \neq \bar{M}_{\Lambda}$, the above expression does not vanish in an obvious way. Nevertheless, one can show that it is in fact zero by using a couple of properties of special geometry. For that purpose, let us rewrite (\ref{DZ}) as:
\be \label{DZp}
D_{\bar{i}} \bar{Z} = -i \,Z \,\bar{L}^{\Lambda} (\bar{{\cal N}}_{\Lambda \Sigma} - {\cal N}_{\Lambda \Sigma}) \,D_{\bar{i}} \bar{L}^{\Sigma} \, .
\ee
Now, from (\ref{NG}) and its complex conjugate, we have:
\be \label{NbN}
\bar{{\cal N}}_{\Lambda \Sigma} - {\cal N}_{\Lambda \Sigma} = G_{\Lambda \Sigma} - \bar{G}_{\Lambda \Sigma} - 2i \frac{({\rm Im} G_{\Lambda \Gamma}) ({\rm Im} G_{\Sigma \Pi})}{({\rm Im} G_{\Xi \Omega})} \!\left( \frac{\bar{L}^{\Gamma} \bar{L}^{\Pi}}{\bar{L}^{\Xi} \bar{L}^{\Omega}} + \frac{L^{\Gamma} L^{\Pi}}{L^{\Xi} L^{\Omega}} \right).
\ee
Substituting this in (\ref{DZp}) and using that $({\rm Im} G_{\Sigma \Pi}) L^{\Pi} D_{\bar{i}} \bar{L}^{\Sigma} = 0$, see (\ref{ImGLDL}), we find that the $(L^{\Gamma} L^{\Pi})/(L^{\Xi} L^{\Omega})$ term drops out and the remaining terms give:
\bea \label{DZpp}
D_{\bar{i}} \bar{Z} \!\!&=& \!\!-i Z \,D_{\bar{i}} \bar{L}^{\Sigma} \times \\
\!\!&\times& \!\!\frac{2i}{({\rm Im} G_{\Xi \Omega}) \bar{L}^{\Xi} \bar{L}^{\Omega}} \left( ({\rm Im} G_{\Lambda \Sigma}) \bar{L}^{\Lambda} ({\rm Im} G_{\Xi \Omega}) \bar{L}^{\Xi} \bar{L}^{\Omega} - ({\rm Im} G_{\Lambda \Gamma}) \bar{L}^{\Lambda} \bar{L}^{\Gamma} ({\rm Im} G_{\Sigma \Pi}) \bar{L}^{\Pi} \right)\equiv 0 \, . \nn
\eea
Note that, whereas the presence of $D_{\bar{i}} \bar{L}^{\Sigma}$ in (\ref{DZp}) was essential for the vanishing of the $(L^{\Gamma} L^{\Pi})/(L^{\Xi} L^{\Omega})$ term, the cancellation in (\ref{DZpp}) works because of the $\bar{L}^{\Lambda}$ multiplier. For future use, let us extract explicitly the special geometry property that follows from the considerations (\ref{DZ})-(\ref{DZpp}):
\be \label{SpRel}
M_{\Lambda} D_i L^{\Lambda} - L^{\Lambda} D_i M_{\Lambda} = L^{\Lambda} ({\cal N}_{\Lambda \Sigma} - \bar{{\cal N}}_{\Lambda \Sigma}) D_i L^{\Sigma} = 0 \, .
\ee

Now let us go back to the generalized attractor equations for flux vacua (\ref{AE2}). We will show that they imply $D_i Z = 0$ in a manner similar to the one in the previous paragraph. From (\ref{Znew}), we have:
\be
D_i Z = q_{fa} D_i L^a - i q_{ha} D_i \tilde{L}^a - p_f^a D_i M_a + i p_h^a D_i \tilde{M}_a \, .
\ee
Substituting the charges with their corresponding expressions in (\ref{AE2}), we find:
\bea \label{DZb}
D_i Z &=& (\tau \bar{Z} M_a + \bar{\tau} Z \bar{M}_a) \,D_i L^a - (\tau \bar{Z} L^a + \bar{\tau} Z \bar{L}^a) \,D_i M_a \\
&-& i (\bar{Z} M_a + Z \bar{M}_a) \,D_i \tilde{L}^a + i (\bar{Z} L^a + Z \bar{L}^a) \,D_i \tilde{M}_a \nn \\
&+& (\bar{\tau} \bar{Z}^{\underline{0}I} D_I M_a + \tau Z^{\underline{0}I} \bar{D}_I \bar{M}_a) \,D_i L^a - (\bar{\tau} \bar{Z}^{\underline{0}I} D_I L^a + \tau Z^{\underline{0}I} \bar{D}_I \bar{L}^a) \,D_i M_a \nn \\
&-& i (\bar{Z}^{\underline{0}I} D_I M_a + Z^{\underline{0}I} \bar{D}_I \bar{M}_a) \,D_i \tilde{L}^a + i (\bar{Z}^{\underline{0}I} D_I L^a + Z^{\underline{0}I} \bar{D}_I \bar{L}^a) \,D_i \tilde{M}_a \, , \nn
\eea
where we have rearranged the various terms in a convenient way. One can immediately notice that the first line vanishes for exactly the same reasons as in the BH case. Namely, due to $D_i M_a = \bar{{\cal N}}_{ab} D_i L^b$ and $M_a = {\cal N}_{ab} L^b$, the $Z$ terms cancel out right away\footnote{In (\ref{DZp}) the $\bar{Z}$ terms were canceling for that reason; the difference is obviously due to the fact that there we were considering $D_{\bar{i}} \bar{Z}$ instead of $D_i Z$.}, whereas the proof that the $\bar{Z}$ terms vanish follows exactly the arguments leading to (\ref{SpRel}). Now let us turn to the second line in (\ref{DZb}). Since $D_i \tilde{M}_a = -i \tau D_i M_a$ and $D_i \tilde{L}^a = -i \tau D_i L^a$, it is clear that this line too gives zero in precisely the same way as for the BH case in the previous paragraph. Of course, it is not surprising that so far nothing new was needed since the first two lines of (\ref{DZb}) come from the attractor equations (\ref{AE1}) for the $D_A D_B Z = 0$ case, which is related to the BH attractors in a very simple way as we saw in the beginning of Subsection \ref{NewAttrSimple}. The new ingredients in the present case are the last two lines in (\ref{DZb}). So one might expect that some additional special geometry properties may be needed in order to verify their vanishing. Indeed, if one wants to follow the same logic as for the first two lines, namely to cancel the terms on the third line among themselves (and the same for the terms on the fourth  line), one runs into a problem. More precisely, whereas the $\bar{Z}^{\underline{0}I}$ terms cancel out right away\footnote{This time one only needs to use $D_i M_a = \bar{{\cal N}}_{ab} D_i L^b$ in both terms (together with $\bar{{\cal N}}_{ab} = \bar{{\cal N}}_{ba}$).}, the $Z^{\underline{0}I}$ terms give:
\be
\tau Z^{\underline{0}\bar{j}} \,D_{\bar{j}} \bar{L}^b ({\cal N}_{ab} - \bar{\cal{N}}_{ab}) D_i L^a = -i \tau Z^{\underline{0}\bar{j}} g_{i\bar{j}} \neq 0 \, ,
\ee
where we have used equation (3.10) in \cite{CDFvP}. This result may look worrisome. However, interestingly enough things work out in a much simpler way. Namely, the $Z^{\underline{0}I}$ terms on the third line cancel in a straightforward manner with the $Z^{\underline{0}I}$ terms on the fourth line due to $D_i \tilde{L}^a = -i \tau D_i L^a$ and $D_i \tilde{M}_a = -i \tau D_i M_a$. Finally, the $\bar{Z}^{\underline{0}I}$ terms on the fourth line cancel each other similarly to the $\bar{Z}^{\underline{0}I}$ terms in the third line. So we have shown that when the new attractor equations (\ref{AE2}) are satisfied, then one also has that $D_i Z = 0$.

To complete the proof that (\ref{AE2}) give solutions to the supersymmetry conditions, we also need to verify that they imply $D_{\tau} Z = 0$. For that purpose, we again consider (\ref{DtauZ}):
\be
D_{\tau} Z = q_{fa} D_{\tau} L^a - i q_{ha} D_{\tau} \tilde{L}^a -p_f^a D_{\tau} M_a + ip_h^a D_{\tau} \tilde{M}_a \, .
\ee
As before, we will substitute here the charges with their corresponding expressions from (\ref{AE2}). Recall that we have already shown the vanishing of all terms that do not contain $Z^{\underline{0}I}$ or $\bar{Z}^{\underline{0}I}$ (see equations (\ref{Dtau})-(\ref{Ktau}) and the discussion around them). So we are left with:
\bea \label{Dtaup}
D_{\tau} Z \!\!\!&=& \!\!\!(\bar{\tau} \bar{Z}^{\underline{0}I} D_I M_a + \tau Z^{\underline{0}I} \bar{D}_I \bar{M}_a) K_{\tau} L^a - (\bar{Z}^{\underline{0}I} D_I M_a + Z^{\underline{0}I} \bar{D}_I \bar{M}_a) (1 + \tau K_{\tau}) L^a \nn \\
\!\!\!&-& \!\!\!(\bar{\tau} \bar{Z}^{\underline{0}I} D_I L^a + \tau Z^{\underline{0}I} \bar{D}_I \bar{L}^a) K_{\tau} M_a + (\bar{Z}^{\underline{0}I} D_I L^a + Z^{\underline{0}I} \bar{D}_I \bar{L}^a) (1 + \tau K_{\tau}) M_a \nn \\
\!\!\!&=& \!\!\!Z^{\underline{0}I} \bar{D}_I (\bar{L}^a M_a - L^a \bar{M}_a) + \bar{Z}^{\underline{0}I} (M_a D_I L^a - L^a D_I M_a) (1 + (\tau - \bar{\tau}) K_{\tau}) \, .
\eea 
Clearly the $Z^{\underline{0}I}$ term vanishes due to $ D_i (\bar{L}^a M_a - L^a \bar{M}_a) = 0$\,, whereas the $\bar{Z}^{\underline{0}I}$ term is zero because of either (\ref{Ktau}) or (\ref{SpRel}). Thus, we conclude that indeed equations (\ref{AE2}) lead to $D_{\tau} Z = 0$. Note that, unlike (\ref{Dtau}), the vanishing of (\ref{Dtaup}) follows entirely from the special K\"{a}hler geometry of the complex structure moduli, regardless of the form of the K\"{a}hler potential for $\tau$.

In this section we showed that the new attractor equations (\ref{AE2}) automatically give solutions to the susy conditions. In principle, one also needs to verify that they are compatible with (\ref{Znew}), i.e. that the right-hand side of (\ref{Znew}), with charges substituted from their attractor expressions, gives exactly $Z$. This is not expected to be an issue for the concrete case of (\ref{AE2}) though, since $\int_{X_4} F_4 \wedge \hat{\Omega}_4 = Z \int_{X_4} \bar{\hat{\Omega}}_4 \wedge \hat{\Omega}_4 = Z$ with $X_4 = (CY(3)\times T^2)/ \bb{Z}_2$\,, as noted in \cite{RK}. Nevertheless, for completeness we explain in Appendix \ref{app} why in general $DZ = 0$ can have spurious solutions and go on to show explicitly that the new attractors (\ref{AE2}) do indeed satisfy (\ref{Znew}).

\section{Generalizations} \label{Gen}
\setcounter{equation}{0}

It is natural to ask whether the flux vacua attractors of \cite{RK} are a peculiarity of type IIB CY orientifolds with flux or whether they are a special case of a more general picture. This question is of significant importance, since if the attractor mechanism could be extended to generic $N=1$ flux vacua, that would provide a new perspective on moduli stabilization. With this motivation in mind, let us now explore the various possibilities for generalizing Kallosh's new attractor equations.

\subsection{Heterotic on non-K\"{a}hler manifolds} \label{HetnonKah}

A first possibility for generalization, addressed in \cite{GD}, is to consider the heterotic string on non-K\"{a}hler manifolds. Before scrutinizing the details of this proposal, let us first review some basic properties of $SU(3)$ structure manifolds and recall the arguments of \cite{GD} in favour of this avenue for generalization of the flux vacua attractors. 

\subsubsection{$SU(3)$ structure compactifications}

Generically, heterotic string compactifications with background fluxes require the internal manifold to have $SU(3)$ structure, instead of $SU(3)$ holonomy. Such manifolds are still characterized by the existence of an almost complex structure $J$ and a holomorphic three-form $\Omega$. However, unlike for a CY, $J$ and $\Omega$ are not closed with respect to the Levi-Civita connection. Their non-closedness is encoded in five torsion classes; for more details see \cite{CCDLMZ}. What is important for us is that the complex and K\"{a}hler structure moduli spaces are special K\"{a}hler manifolds with K\"{a}hler potentials \cite{GLW1}:
\bea \label{KJKOm}
K_J &=& -\ln i \int \la e^{-J_c} , e^{-\bar{J}_c} \ra = -\ln \frac{4}{3} \int J\wedge J\wedge J \,\, , \nn \\
K_{\Omega} &=& -\ln i \int \la \Omega , \bar{\Omega} \ra = -\ln i \int \Omega \wedge \bar{\Omega} \,\, ,
\eea
where $J_c = B + iJ$ and the bracket $\la \,, \ra$ denotes the Mukai pairing, defined by:
\bea \label{Mukai}
\la \varphi , \psi \ra &=& - \varphi_1 \wedge \psi_5 + \varphi_3 \wedge \psi_3 - \varphi_5 \wedge \psi_1 \qquad \qquad \quad \,\,\,\, {\rm for\,\, odd \,\, forms} \nn \\
\la \varphi , \psi \ra &=& \varphi_0 \wedge \psi_6 - \varphi_2 \wedge \psi_4 + \varphi_4 \wedge \psi_2 - \varphi_6 \wedge \psi_0 \qquad {\rm for\,\, even \,\, forms}  
\eea 
with $\varphi_p$ being the $p$-form component of the mixed-degree form $\varphi$ and similarly for $\psi$. The moduli of the compactification arise from the expansion of $e^{-J_c}$ and $\Omega$ in terms of a basis of forms \cite{GLW1}:
\bea \label{Expand}
e^{-J_c} &=& X^0 (t) + X^{\alpha}(t) \,\omega_{\alpha} - G_{\alpha} (t) \,\tilde{\omega}^{\alpha} - G_0 (t) \star 1 \, , \nn \\
\Omega &=& X^{\cal I} (z) \,\alpha_{\cal I} - G_{\cal I} (z) \,\beta^{\cal I} \, ,
\eea
where $t^{\alpha}$ and $z^i$ denote the K\"{a}hler and complex structure moduli respectively, $(\alpha_{\cal I} , \beta^{\cal I})$ are a basis for the 3-forms and $1, \omega_{\alpha}, \tilde{\omega}^{\alpha}, \star 1$ are a basis for the 0-, 2-, 4- and 6-forms.\footnote{It is beneficial to consider $e^{-J_c}$, instead of just $J_c$, in order to easily view $SU(3)$ structure manifolds as a special case of the more general $SU(3)\times SU(3)$ structure spaces. Recall that the latter are most naturally described in terms of a pair of pure spinors $(\Phi_+ , \Phi_-)$, which for $SU(3)$ structure reduces to $(e^{-J_c} , \Omega)$. We will give more details on $SU(3)\times SU(3)$ structure spaces in the next subsection.} Denoting $\omega_{\cal A} \equiv (1, \omega_{\alpha})$ and $\tilde{\omega}^{\cal A} \equiv (\star 1, \tilde{\omega}^{\alpha})$, we can write:\footnote{The index notation in (\ref{Expand}) and (\ref{EvExp}) differs from the one in \cite{GD}; we have adopted it for future convenience, i.e. for easier comparison with the literature on $SU(3)\times SU(3)$ structure compactifications that we will be considering in Subsection \ref{SU3SU3}.}
\be \label{EvExp}
e^{-J_c} = X^{\cal A} (t) \,\omega_{\cal A} - G_{\cal A} (t) \,\tilde{\omega}^{\cal A} \, .
\ee

The basis forms introduced above satisfy:
\bea \label{symppair}
\int \la \alpha_{\cal I} , \beta^{\cal J} \ra = - \int \la \beta^{\cal J} , \alpha_{\cal I} \ra = \delta_{\cal I}^{\cal J} &,& \quad \int \la \alpha_{\cal I} , \alpha_{\cal J} \ra = 0 = \int \la \beta^{\cal I} , \beta^{\cal J} \ra \,\,\,\, , \nn \\
\int \la \omega_{\cal A} , \tilde{\omega}^{\cal B} \ra = - \int \la \tilde{\omega}^{\cal B} , \omega_{\cal A} \ra = \delta_{\cal A}^{\cal B} &,& \quad \int \la \omega_{\cal A} , \omega_{\cal B} \ra = 0 = \int \la \tilde{\omega}^{\cal A} , \tilde{\omega}^{\cal B} \ra \,\,\,\, ,
\eea
where the integration is over the internal manifold and the Mukai pairing $\la \,, \ra$ was defined in (\ref{Mukai}). Also, the pairing of any even form with any odd form gives zero. For a generic $SU(3)$ structure manifold, these basis forms are not harmonic. Instead, they satisfy the following differential conditions \cite{GLW1,dCGLM}:
\be \label{HaFl}
d \omega_{\alpha} = m_{\alpha}^{\cal I} \alpha_{\cal I} - e_{{\cal I} \alpha} \beta^{\cal I} \, , \qquad d \tilde{\omega}^{\alpha} = 0 \, , \qquad d \alpha_{\cal I} = e_{{\cal I} \alpha} \tilde{\omega}^{\alpha} \, , \qquad d \beta^{\cal I} = m_{\alpha}^{\cal I} \tilde{\omega}^{\alpha} \, ,
\ee
where the constant matrices $m_{\alpha}^{\cal I}$ and $e_{\alpha {\cal I}}$ are constrained by
\be \label{NilConstr}
m_{\alpha}^{\cal I} e _{{\cal I} \beta} - e_{{\cal I} \alpha} m_{\beta}^{\cal I} = 0 
\ee
in order to ensure the nilpotency of the exterior differential. We should note that the constraint (\ref{NilConstr}) is relevant for the standard embedding. For nonstandard embeddings, the BI of the NS 3-form flux $H$ is non-trivial and this may result in other restrictions on the fluxes/charges \cite{dCGLM}. As in \cite{GD}, we will not consider the gauge moduli and will only concentrate on the geometric ones here.

Now, recall that the superpotential in the heterotic case is \cite{HetW}:
\be \label{HetSupP}
W = \int (H + dJ_c) \wedge \Omega \,\, .
\ee
Comparing (\ref{Expand}) with 
\be
e^{-J_c} = 1 - J_c +\frac{1}{2} J_c\wedge J_c - \frac{1}{6} J_c \wedge J_c \wedge J_c \,\, ,
\ee
we can see that $X^0 = 1$ and $J_c = - X^{\alpha} (t) \omega_{\alpha}$. Taking, as usual, the special coordinates to be $t^{\alpha} = X^{\alpha}/X^0$, we then have $J_c = - t^{\alpha} \omega_{\alpha}$. Using this, together with (\ref{HaFl}) and the decomposition of the NS flux as
\be
H = p^{\cal I} \alpha_{\cal I} - q_{\cal I} \beta^{\cal I} \, ,
\ee
one finds that the superpotenial (\ref{HetSupP}) acquires the following form \cite{GD}:
\be \label{hetW}
W = (q_{\cal I} - t^{\alpha} e_{{\cal I} \alpha}) \,X^{\cal I} (z) - (p^{\cal I} - t^{\alpha} m_{\alpha}^{\cal I} ) \,G_{\cal I} (z) \,\, .
\ee
This expression resembles a lot the superpotential (\ref{Sup}) for the case of IIB CY orientifolds, considered in \cite{RK}. The similarity with the IIB case is even more apparent by rewriting (\ref{hetW}) as:
\be
W = \int (H - t^{\alpha} F_{\alpha}) \wedge \Omega \, ,
\ee
where $F_{\alpha} \equiv d\omega_{\alpha}$. Specializing to the case of a single K\"{a}hler modulus $t$, one has $W_{het} = \int (H - t F) \wedge \Omega$, which looks exactly like the type IIB superpotential $W_{IIB} = \int (F - \tau H)\wedge \Omega$. This, together with the fact that the complex structure moduli space in both cases has special K\"{a}hler geometry, inspired the author of \cite{GD} to make a particular conjecture for heterotic attractors. Before considering an arbitrary number of K\"{a}hler moduli, let us first examine more carefully this single-modulus conjecture.

\subsubsection{Single K\"{a}hler modulus} \label{SingleKahMod}

The conjecture of \cite{GD} is that for the case of a single K\"{a}hler modulus there are supersymmetric heterotic attractor equations of exactly the same form as (\ref{AE2}). The obvious substitutions, that one has to make in the latter equations in order to obtain the heterotic ones, are: $\tau \rightarrow t$ and $(p_h^a, q_{ha}, p_f^a, q_{fa})\rightarrow (m^{\cal I},e_{\cal I}, p^{\cal I}, q_{\cal I})$. For completeness, let us write down the proposed attractor equations in the more concise manner used in \cite{GD}. Introducing the following notation: $L^{\cal I}_{0} = e^{(K_J + K_{\Omega})/2} X^{\cal I}$ and $M_{\cal I}^0 = e^{(K_J + K_{\Omega})/2} G_{\cal I}$\,, together with ${\cal V} = (L^{\cal I}_0,M_{\cal I}^0)$, ${\cal Q}_F = (m^{\cal I},e_{\cal I})$ and ${\cal Q}_H = (p^{\cal I}, q_{\cal I})$, they acquire the form:\footnote{In \cite{GD}, the proposed conjecture differs from (\ref{PrHetAttr}) by having $(K_t)^{-1}$ instead of the vielbein $e^t_{\underline{t}}$. For the original new attractors of \cite{RK}, one has $(K_{\tau})^{-1} = e^{\tau}_{\underline{\tau}}$ due to the specific form of the K\"{a}hler potential for $\tau$ in that case. In general, however, $(K_t)^{-1}\neq e^t_{\underline{t}}$. In any case, our subsequent arguments are independent of which one of those two options is taken in (\ref{PrHetAttr}).}
\be \label{PrHetAttr}
\left( 
\begin{array}{c}
{\cal Q}_F \\
{\cal Q}_H \\
\end{array}
\right) = 
\left(
\begin{array}{c}
2 {\rm Re} (\bar{Z} {\cal V}) \\
2 {\rm Re} (t \bar{Z} {\cal V}) \\
\end{array}
\right) +
\left(
\begin{array}{c}
2 {\rm Re} (g^{i\bar{j}} \,e^{t}_{\underline{t}} \, D_{t} D_{i} Z \,\bar{D}_{\bar{j}} \bar{{\cal V}}) \\
2 {\rm Re} (t \,g^{i\bar{j}} \,e^{t}_{\underline{t}} \, D_{t} D_{i} Z \,\bar{D}_{\bar{j}} \bar{{\cal V}}) \\
\end{array}
\right) .
\ee

To verify this conjecture, one has to check that the proposed attractor equations (\ref{PrHetAttr}) imply the supersymmetry conditions $D_{z^i} Z = 0$ and $D_t Z = 0$. This computation follows closely our considerations in Section \ref{AttrProof} regarding the attractors of \cite{RK}. From those considerations, it is immediately clear that the form of the first derivative of the K\"{a}hler potential with respect to $t$ is of crucial importance. More precisely, while the vanishing of $D_i Z$ is due to the special K\"{a}hler geometry of the complex structure moduli exactly as in Section {\ref{AttrProof}}, the $D_t Z$ derivative vanishes only if
\be \label{KC}
1 - \bar{t} K_t + t K_t = 0 \, ,
\ee
as in (\ref{Dtau}). Now, it is easy to convince oneself that (\ref{KC}) is, in fact, {\it not} satisfied, the reason being that the expression for $K_t$ is not exactly the same as that for $K_{\tau}$. To be more explicit, let us write out the relevant K\"{a}hler potential. For a single K\"{a}hler modulus we have: $J_c = B+iJ = \,-\,t \,\omega$ with $\omega$ a 2-form. Then (\ref{KJKOm}) implies:
\be \label{KJsingle}
K_J = - \ln i \int \la e^{-J_c} , e^{-\bar{J}_c} \ra = - \ln \left[ \frac{i}{6} \left( \bar{t} - t \right)^3 \right] ,
\ee
where we have used that
\be
e^{-J_c} = e^{t \omega} = 1 + t \,\omega + \frac{t^2}{2} \,\omega \wedge \omega + \frac{t^3}{6} \,\omega \wedge \omega \wedge \omega
\ee
and also the definition of the Mukai pairing in (\ref{Mukai}).\footnote{We have also assumed the normalization $\int \omega \wedge \omega \wedge \omega = 1$. Clearly, any other normalization only gives a constant numerical factor in front of the bracket inside the logarithm in (\ref{KJsingle}) and therefore is irrelevant for the computation of any derivatives of $K_J$.} Alternatively, one can compute $K_J$ by substituting $J=\frac{i}{2} (t-\bar{t}) \,\omega$ into $K_J = -\ln \frac{4}{3} \int J\wedge J\wedge J$, finding again (\ref{KJsingle}). So we see that 
\be \label{KJt}
\pd_t K_J = - \frac{3}{t - \bar{t}}
\ee
and hence
\be
1 + (t - \bar{t}) \,\pd_t K_J = -2 \, ,
\ee
which implies that $D_t Z \neq 0$ as explained above. 

It might seem that one could compensate the problematic factor of 3 in (\ref{KJt}) by introducing suitable numerical coefficients in front of the different rows on the right-hand side of (\ref{PrHetAttr}). Unfortunately, however, this is not possible. The reason is that the $1$ and the $t$ term in the expression $1 + (t - \bar{t}) \,\pd_t K_J$ have the same origin, whereas the $\bar{t}$ term has different origin. In other words, if one multiplies the rhs of the ${\cal Q}_F$ rows in (\ref{PrHetAttr}) with a numerical constant $r$ and the rhs of the ${\cal Q}_H$ rows with another constant $s$, then one finds that \,$1 + (t - \bar{t}) \,\pd_t K_J \,\,\rightarrow \,\,r + (r t - s \bar{t}) \,\pd_t K_J$\,. Clearly, no choice of $r$ and $s$ can make the last expression vanish for arbitrary $t$.

An additional problem with the proposal (\ref{PrHetAttr}) is that it is not properly normalized. Indeed, the analogue of the computation in (\ref{ApComp}) does not give $Z$. Instead, one finds that: $(q_{\cal I} - t \,e_{\cal I}) L_0^{\cal I} - (p^{\cal I} - t \,m^{\cal I})M^0_{\cal I} = (6 Z)/(t-\bar{t})^2$. However, recall that at the classical level the heterotic string has only Minkowski susy vacua, unlike type IIB.\footnote{See \cite{FLi} for more details on heterotic AdS vacua in the presence of a gaugino condensate.} And when $Z=0$ the above two problems of (\ref{PrHetAttr}), namely with the normalization and with $D_t Z \neq 0$, actually disappear. So one might be inclined to conclude that the proposal (\ref{PrHetAttr}) works since, although it describes only Minkowski attractors, in the heterotic case this is all that is necessary. We would like to caution, though, that the attractor equations are supposed to originate from an expansion that is valid everywhere in moduli space, not just at the supersymmetric extrema. Hence, the relevant expressions have to be normalized correctly for any $Z\neq 0$, even when they only give vacua for $Z=0$. As a last remark, let us also mention that (\ref{PrHetAttr}) is not of the form that generalizes properly to arbitrary number of K\"{a}hler moduli, as we will see in the following.

In this subsubsection we showed that the single-K\"{a}hler-modulus conjecture (\ref{PrHetAttr}) fails. This does not mean that there are no heterotic attractors, just that one should not treat the K\"{a}hler modulus similarly to the axion-dilaton in (\ref{AE2}). Instead, for the successful formulation of heterotic attractor equations it is essential that the K\"{a}hler and complex structure moduli be treated on equal footing. In order to understand how to do that, it is instructive to consider in more detail the case of several K\"{a}hler moduli. This will also lead us to a technical motivation to look at type II $SU(3)\times SU(3)$ structure compactifications in the search for generalizations of the flux vacua attractors. (The physical motivation, of course, is that they are the most general extension of the situation considered in \cite{RK}, in the realm of geometric compactifications.)

\subsubsection{Heterotic attractors} \label{HetMinkAttr}

Let us now turn to the general case of arbitrary number of K\"{a}hler moduli. It is convenient to introduce the notation: $m_0^{\cal I} = - p^{\cal I}$ and $e_{{\cal I} 0} = - q_{\cal I}$. Then, in terms of $m_{\cal A}^{\cal I} =(m_0^{\cal I}, m_{\alpha}^{\cal I})$ and $e_{\cal I A} = (e_{{\cal I} 0}, e_{{\cal I} \alpha})$, one can write (\ref{hetW}) as:
\be
W = m_{\cal A}^{\cal I} G_{\cal I} (z) X^{\cal A} (t) - e_{\cal I A} X^{\cal I} (z) X^{\cal A} (t) \, ,
\ee
where we have used that $X^0 (t) = 1$ and $X^{\alpha} = t^{\alpha}$. Clearly, it is most natural to define the covariantly holomorphic sections $(L^{\cal I}, M_{\cal I})$ and $(L^{\cal A}, M_{\cal A})$ in the usual way, unlike the sections in (\ref{AE2}) and (\ref{PrHetAttr}), so that the special K\"{a}hler geometries of the complex and K\"{a}hler structure moduli factorize. I.e, we define:
\be
L^{\cal I} = e^{K_{\Omega}/2} X^{\cal I} \, , \quad M_{\cal I} = e^{K_{\Omega}/2} G_{\cal I} \,\, ; \quad L^{\cal A} = e^{K_J/2} X^{\cal A} \, , \quad M_{\cal A} = e^{K_J/2} G_{\cal A} \, .
\ee
In terms of these sections, the central charge becomes:
\be \label{ZLM}
Z = e^{(K_{\Omega}+K_J)/2} W = m_{\cal A}^{\cal I} M_{\cal I} L^{\cal A} - e_{\cal I A} L^{\cal I} L^{\cal A} \, .
\ee
This nice expression suggests that it may be useful to introduce the double-symplectic section \cite{GD}:
\be \label{DoubleSec}
\hat{{\cal V}} = e^{(K_{\Omega}+K_J)/2} (\,\Omega \otimes e^{-J_c}) \, ,
\ee
or in more detail:
\bea \label{Vhat}
\hspace*{-1.3cm}\hat{{\cal V}} &=& (L^{\cal I} \alpha_{\cal I} - M_{\cal I} \beta^{\cal I})\otimes (L^{\cal A} \omega_{\cal A} - M_{\cal A} \tilde{\omega}_{\cal A}) \nn \\
\hspace*{-1.3cm}&=& L^{\cal I} L^{\cal A} \,\alpha_{\cal I} \otimes \omega_{\cal A} - M_{\cal I} L^{\cal A} \,\beta^{\cal I} \otimes \omega_{\cal A} - L^{\cal I} M_{\cal A} \,\alpha_{\cal I} \otimes \tilde{\omega}^{\cal A} + M_{\cal I} M_{\cal A} \,\beta^{\cal I} \otimes \tilde{\omega}^{\cal A} \, .
\eea
Then, using (\ref{symppair}), one can verify that the central charge (\ref{ZLM}) can be written as:
\be \label{ZQV}
Z = - \int \la \hat{{\cal Q}} , \hat{{\cal V}} \ra \, ,
\ee
where $\hat{{\cal Q}} \equiv - F_{\cal A} \otimes \tilde{\omega}^{\cal A}$ with $F_{\cal A} = m_{\cal A}^{\cal I} \alpha_{\cal I} - e_{\cal I A} \beta^{\cal I}$; the minus in the definition of $\hat{{\cal Q}}$ ensures that this generalized "flux" is in the same basis as (\ref{Vhat}). 

Now, it is tempting to propose, similarly to \cite{GD}, that there are supersymmetric attractor equations for heterotic flux vacua of the form:
\be \label{GenKahlConj}
{\hat {\cal Q}} = 2 \,{\rm Re} ( \bar{Z} \hat{{\cal V}} + g^{i \bar{j}} g^{\alpha \bar{\beta}} D_i D_{\alpha} \hat{{\cal V}} \,\bar{D}_{\bar{j}} \bar{D}_{\bar{\beta}} \bar{Z} ) \, .
\ee
Notice that, unlike in \cite{GD}, this proposal has to contain ${\rm Re}$ and not ${\rm Im}$ in order to be consistent with (\ref{ZQV}), since the normalization of the doubled section is:
\be
\int \la \bar{\hat{{\cal V}}}, \hat{{\cal V}} \ra = (\bar{L}^{\cal I} M_{\cal I} - L^{\cal I} \bar{M}_{\cal I}) (\bar{L}^{\cal A} M_{\cal A} - L^{\cal A} \bar{M}_{\cal A}) = (-i)^2 = -1 \, .
\ee
Although at first sight (\ref{GenKahlConj}) might look like a natural generalization of (\ref{PrHetAttr}), it is in fact somewhat different. Indeed, more explicitly it states that:
\bea \label{meNC}
\hspace*{-0.5cm}m_{\cal A}^{\cal I} &=& \bar{Z} L^{\cal I} M_{\cal A} + Z \bar{L}^{\cal I} \bar{M}_{\cal A} + \bar{Z}^{i \alpha} \,D_i L^{\cal I} \,D_{\alpha} M_{\cal A} + Z^{\bar{i} \bar{\alpha}} \,D_{\bar{i}} \bar{L}^{\cal I} \,D_{\bar{\alpha}} \bar{M}_{\cal A} \, , \nn \\ 
\hspace*{-0.5cm}e_{\cal I A} &=& \bar{Z} M_{\cal I} M_{\cal A} + Z \bar{M}_{\cal I} \bar{M}_{\cal A} + \bar{Z}^{i \alpha} \,D_i M_{\cal I} \,D_{\alpha} M_{\cal A} + Z^{\bar{i} \bar{\alpha}} \,D_{\bar{i}} \bar{M}_{\cal I} \,D_{\bar{\alpha}} \bar{M}_{\cal A} \, ,
\eea
where $Z^{\bar{i} \bar{\alpha}} \equiv D^{\bar{i}} D^{\bar{\alpha}} Z = g^{\bar{i} j} g^{\bar{\alpha} \beta} D_j D_{\beta} Z$. These expressions for $m_{\cal A}^{\cal I}$ and $e_{\cal I A}$ {\it do not} reduce to (\ref{PrHetAttr}) when there is only one K\"{a}hler modulus because $(e^{K_J/2}, \,e^{K_J/2} \,t^{\alpha})$ is $L^{\cal A}$, not $M_{\cal A}$, and also because of the minus signs in $m_0^{\cal I} = - p^{\cal I}$ and $e_{{\cal I} 0} = -q_{\cal I}$. This is a good sign, given that the conjecture (\ref{PrHetAttr}) does not work as we saw above. So it is worth exploring the proposal (\ref{meNC}) in more detail. Before turning to that however, let us make an important remark.

By comparing (\ref{Vhat}) with $\hat{{\cal Q}} = - m_{\cal A}^{\cal I} \alpha_{\cal I} \otimes \tilde{\omega}_{\cal A} + e_{\cal I A} \beta^{\cal I} \otimes \tilde{\omega}^{\cal A}$, it is easy to realize that in the case of the heterotic string some of the possible charges are identically zero. The general case corresponds to a generalized flux of the form:
\be \label{Qgen}
\hat{{\cal Q}} = \tilde{m}^{\cal I A} \,\alpha_{\cal I} \otimes \omega_{\cal A} - \tilde{e}_{\cal I}^{\cal A} \,\beta^{\cal I} \otimes \omega_{\cal A} - m^{\cal I}_{\cal A} \,\alpha_{\cal I} \otimes \tilde{\omega}^{\cal A} + e_{\cal I A} \,\beta^{\cal I} \otimes \tilde{\omega}^{\cal A} \, ,
\ee 
or in other words to the charge matrix
\be \label{Qtot}
{\cal Q} = \left( 
\begin{array}{cc}
\tilde{m}^{{\cal I} {\cal A}} & m_{\cal A}^{\cal I} \\
\tilde{e}_{\cal I}^{\cal A} & e_{{\cal I} {\cal A}} \\
\end{array}
\right) 
\ee
that appears in type II $SU(3)\times SU(3)$ structure compactifications \cite{GLW2,CB} (it was also introduced in \cite{GD}).\footnote{The meaning of the charge matrix $Q$ will become more clear in Section \ref{SU3SU3} and, in particular, in equation (\ref{DiffCond}).} In such compactifications all elements of ${\cal Q}$ can be non-vanishing, unlike the $SU(3)$ structure case in which $\tilde{m}^{\cal I A} \equiv 0 \equiv \tilde{e}^{\cal A}_{\cal I}$.

Let us now go back to the proposal (\ref{meNC}). Substituting it in $D_j Z$\,, we find:
\bea \label{Dj}
D_j Z &=& m_{\cal A}^{\cal I} (D_j M_{\cal I}) L^{\cal A} - e_{\cal I A} (D_j L^{\cal I}) L^{\cal A} \nn \\ 
&=& \bar{Z} \,M_{\cal A} L^{\cal A} \,L^{\cal I} (\bar{{\cal N}}_{{\cal I J}} - {\cal N}_{{\cal I J}}) D_j L^{\cal J} + Z \, \bar{M}_{\cal A} L^{\cal A} \,D_j (\bar{L}^{\cal I} M_{\cal I} - L^{\cal I} \bar{M}_{\cal I}) \nn \\ 
&+& \bar{Z}^{i \alpha} \,L^{\cal A} \,(D_{\alpha} M_{\cal A}) \,(D_i L^{\cal I} \bar{{\cal N}}_{\cal I J} D_j L^{\cal J} - D_i L^{\cal J} \bar{{\cal N}}_{\cal I J} D_j L^{\cal I}) \nn \\
&+& Z^{\bar{i} \bar{\alpha}} \,L^{\cal A} \,(D_{\bar{\alpha}} \bar{M}_{\cal A}) \,D_{\bar{i}} (\bar{L}^{\cal I} D_j M_{\cal I} - \bar{M}_{\cal I} D_j L^{\cal I}) = \,0 \, , 
\eea
where each of the $Z$, $\bar{Z}$, $Z^{\bar{i} \bar{\alpha}}$ and $\bar{Z}^{i \alpha}$ terms vanishes separately due to the special K\"{a}hler geometry of the complex structure moduli; see (\ref{Norm}) and (\ref{SpRel}) and also recall that ${\cal N}_{\cal I J} = {\cal N}_{\cal J I}$. On the other hand, the derivative with respect to the K\"{a}hler moduli gives:
\bea \label{Dalpha}
D_{\beta} Z &=& (m_{\cal A}^{\cal I} M_{\cal I} - e_{{\cal I A}} L^{\cal I}) D_{\beta} L^{\cal A} = Z \bar{M}_{\cal A} (\bar{L}^{\cal I} M_{\cal I} - L^{\cal I} \bar{M}_{\cal I}) D_{\beta} L^{\cal A} \nn \\
&+& \bar{Z}^{i \alpha} (D_{\alpha} M_{\cal A}) (D_{\beta} L^{\cal A}) (M_{\cal I} D_i L^{\cal I} - L^{\cal I} D_i M_{\cal I}) \nn \\
&+& Z^{\bar{i} \bar{\alpha}} (D_{\bar{\alpha}} \bar{M}_{\cal A}) (D_{\beta} L^{\cal A}) \,D_{\bar{i}} (\bar{L}^{\cal I} M_{\cal I} - L^{\cal I} \bar{M}_{\cal I}) = -i Z \bar{M}_{\cal A} D_{\beta} L^{\cal A} \, ,
\eea
where again the $\bar{Z}^{i \alpha}$ and $Z^{\bar{i} \bar{\alpha}}$ terms vanish because of special K\"{a}hler geometry. Although the final result in (\ref{Dalpha}) is nonzero for a generic $Z$, recall that classically the heterotic string only has Minkowski susy vacua. Nevertheless, it is interesting to note that we do not have to take $Z=0$ by hand in order to ensure $D_{\beta} Z = 0$; the vanishing of $Z$ follows from the attractor proposal (\ref{GenKahlConj}) itself, as we will show below. The reason is that (\ref{GenKahlConj}) contains also the components $\tilde{m}^{\cal I A} \equiv 0 \equiv \tilde{e}_{\cal I}^{\cal A}$\,, which impose certain constraints on the moduli. Note however, that it is a nontrivial statement that the consequence of those constraints is the vanishing of the central charge. In principle, it could have been conceivable that those constraints lead to the vanishing of $Z \bar{M}_{\cal A} D_{\beta} L^{\cal A}$ with $Z\neq 0$.

To illustrate the last point, let us again look at the example of the BH attractors that were reviewed in Subsection \ref{BHat}. The heterotic case is analogous to taking, say, $p^{\Lambda} \equiv 0$ in (\ref{BHCC}) and (\ref{AttrEqBH}). In other words, one has the central charge $Z=q_{\Lambda} L^{\Lambda}$ and the attractor equations $q_{\Lambda} = i (\bar{Z} M_{\Lambda} - Z \bar{M}_{\Lambda})$, together with the constraints
\be \label{constrBH}
i (\bar{Z} L^{\Lambda} - Z \bar{L}^{\Lambda}) = 0 \, .
\ee
So one might worry that now $D_i Z = q_{\Lambda} D_i L^{\Lambda}$ is non-vanishing since the term $p^{\Lambda} D_i M_{\Lambda}$ was essential for the necessary cancellations; see (\ref{DZ})-(\ref{SpRel}). However, the constraints (\ref{constrBH}) imply that:
\be
\bar{Z} X^{\Lambda} - Z \bar{X}^{\Lambda} = 0\, ,
\ee
where we have used $e^{K/2} \neq 0$. Acting with $D_i$ on the last equation, we find:
\be
\bar{Z} \,D_i X^{\Lambda} = \bar{X}^{\Lambda} D_i Z \, .
\ee
Now, in special coordinates $X^0 = 1$, $X^j = t^j$ and so $D_i X^j = \pd_i X^j = \delta_i^j$ whereas $D_i X^0 = 0$. Therefore, taking $\Lambda = 0$ in the last equation, we conclude that $D_i Z = 0$. Still, one may be worried that in the present case we could be forced to have $Z = 0$, since both the $q_{\Lambda}$ and $p^{\Lambda}$ terms are needed in order to verify that the expression $q_{\Lambda} L^{\Lambda} - p^{\Lambda} M_{\Lambda}$, with attractor equations substituted, does equal to $Z$. Indeed, for $p^{\Lambda}=0$ we have 
\be
q_{\Lambda} L^{\Lambda} = i (\bar{Z} M_{\Lambda} L^{\Lambda} - Z \bar{M}_{\Lambda} L^{\Lambda}) \, , 
\ee
which seems quite different from $Z$. However, using the constraints (\ref{constrBH}) to express $\bar{Z} L^{\Lambda}$ as $Z \bar{L}^\Lambda$, we find that 
\be
q_{\Lambda} L^{\Lambda} = i Z (\bar{L}^{\Lambda} M_{\Lambda} - L^{\Lambda} \bar{M}_{\Lambda}) = Z \, .
\ee
Hence a set of charges such that $q_{\Lambda} \neq 0$ and $p^{\Lambda} = 0$ leads to $D_i Z = 0$ with $Z\neq 0$, just like in the general case of BH attractors. 

Let us now get back to the heterotic constraints $\tilde{m}^{\cal I A} \equiv 0$\,, \,$\tilde{e}_{\cal I}^{\cal A} \equiv 0$ and explore their consequences for (\ref{Dalpha}) and, in particular, for the central charge (\ref{ZLM}). We will see that things are somewhat different compared to the BH case of the previous paragraph. To elaborate on that, let us first write down explicitly the constraints that follow from the $\tilde{m}^{\cal I A}$ and $\tilde{e}_{\cal I}^{\cal A}$ components of (\ref{GenKahlConj}): 
\bea \label{constrHet}
0 &=& \bar{Z} L^{\cal I} L^{\cal A} + Z \bar{L}^{\cal I} \bar{L}^{\cal A} + \bar{Z}^{i \alpha} \,D_i L^{\cal I} \,D_{\alpha} L^{\cal A} + Z^{\bar{i} \bar{\alpha}} \,D_{\bar{i}} \bar{L}^{\cal I} \,D_{\bar{\alpha}} \bar{L}^{\cal A} \, , \nn \\
0 &=& \bar{Z} M_{\cal I} L^{\cal A} + Z \bar{M}_{\cal I} \bar{L}^{\cal A} + \bar{Z}^{i \alpha} \,D_i M_{\cal I} \,D_{\alpha} L^{\cal A} + Z^{\bar{i} \bar{\alpha}} \,D_{\bar{i}} \bar{M}_{\cal I} \,D_{\bar{\alpha}} \bar{L}^{\cal A} \, .
\eea
Now, let us substitute the attractor expressions (\ref{meNC}) into $m_{\cal A}^{\cal I} M_{\cal I} L^{\cal A} - e_{\cal I A} L^{\cal I} L^{\cal A}$ in order to see whether the last expression gives $Z$ once the constraints (\ref{constrHet}) are taken into account, similarly to the BH attractor case. From (\ref{meNC}), we have:
\be \label{Check}
m_{\cal A}^{\cal I} M_{\cal I} L^{\cal A} - e_{\cal I A} L^{\cal I} L^{\cal A} = -i Z \bar{M}_{\cal A} L^{\cal A} \, ,
\ee
where again we have used the special geometry properties that $i (\bar{L}^{\cal I} M_{\cal I} - L^{\cal I} \bar{M}_{\cal I}) = 1$ and $M_{\cal I} (D_i L^{\cal I}) - L^{\cal I} (D_i M_{\cal I}) = 0$; see (\ref{SpRel}). Clearly, for $Z\neq 0$ the right-hand side of (\ref{Check}) could equal $Z$ only if it were true that $\bar{M}_{\cal A} L^{\cal A} = i$, which is incompatible with the normalization $i (\bar{L}^{\cal A} M_{\cal A} - L^{\cal A} \bar{M}_{\cal A}) = 1$. However, we will show now that the constraints (\ref{constrHet}) enforce precisely the vanishing of the central charge. Indeed, let us consider the expression:
\be \label{ExprZero}
\bar{Z} L^{\cal I} L^{\cal A} M_{\cal A} M_{\cal I} - \bar{Z} M_{\cal I} L^{\cal A} M_{\cal A} L^{\cal I} + \bar{Z}^{i \alpha} \,D_i L^{\cal I} \,D_{\alpha} L^{\cal A} M_{\cal A} M_{\cal I} - \bar{Z}^{i \alpha} \,D_i M_{\cal I} \,D_{\alpha} L^{\cal A} M_{\cal A} L^{\cal I} \, ,
\ee 
which is identically zero since the first two terms obviously cancel each other whereas the last two cancel due to special geometry. We will use (\ref{constrHet}) to substitute all $\bar{Z}$, $\bar{Z}^{i \alpha}$ terms in (\ref{ExprZero}) with $Z$, $Z^{\bar{i} \bar{\alpha}}$ terms. To do that, let us multiply the first condition in (\ref{constrHet}) by $M_{\cal A} M_{\cal I}$ and the second one by $M_{\cal A} L^{\cal I}$. Then we find that (\ref{ExprZero}) acquires the form:
\be
0 = i Z \,\bar{L}^{\cal A} M_{\cal A} \, ,
\ee
where the $Z^{\bar{i} \bar{\alpha}}$ terms again canceled due to special geometry. Since the normalization of the $(L^{\cal A} , M_{\cal A})$ section implies that $\bar{L}^{\cal A} M_{\cal A} \neq 0$, we conclude that $Z = 0$ as a result of the constraints $\tilde{m}^{\cal I A} \equiv 0 \equiv \tilde{e}_{\cal I}^{\cal A}$. Clearly then, $D_{\beta} Z$ also vanishes as a consequence of the heterotic constraints. To recapitulate, we have established that (\ref{GenKahlConj}) gives heterotic susy attractors for Minkowski vacua.

Before concluding this subsection, let us make a final remark. From the form of the result in (\ref{Dalpha}), it is evident that $D_{\beta} Z$ could be made to vanish, without restricting the value of $Z$, if there were an additional $i Z \,\bar{L}^{\cal A} \,D_{\beta} M_{\cal A}$ term, so that the combined result would be \,$i\,Z \,D_{\beta} (\bar{L}^{\cal A} M_{\cal A} - L^{\cal A} \bar{M}_{\cal A})$\,. It is easy to realize that this would be the case for a theory with central charge of the form (up to a constant):
\be \label{ZSU3sq}
Z = \int \la \hat{{\cal Q}} , \hat{{\cal V}} \ra = e_{{\cal I} {\cal A}} L^{\cal I} L^{\cal A} - m_{\cal A}^{\cal I} M_{\cal I} L^{\cal A} - \tilde{e}_{\cal I}^{\cal A} L^{\cal I} M_{\cal A} + \tilde{m}^{{\cal I} \cal A} M_{\cal I} M_{\cal A} \,\, .
\ee
Indeed, in such a case one can easily verify that the attractor equations (\ref{meNC}) together with the remaining content of (\ref{GenKahlConj}), i.e.
\bea
\tilde{m}^{\cal I A} &=& \bar{Z} L^{\cal I} L^{\cal A} + Z \bar{L}^{\cal I} \bar{L}^{\cal A} + \bar{Z}^{i \alpha} \,D_i L^{\cal I} \,D_{\alpha} L^{\cal A} + Z^{\bar{i} \bar{\alpha}} \,D_{\bar{i}} \bar{L}^{\cal I} \,D_{\bar{\alpha}} \bar{L}^{\cal A} \nn \\
\tilde{e}_{\cal I}^{\cal A} &=& \bar{Z} M_{\cal I} L^{\cal A} + Z \bar{M}_{\cal I} \bar{L}^{\cal A} + \bar{Z}^{i \alpha} \,D_i M_{\cal I} \,D_{\alpha} L^{\cal A} + Z^{\bar{i} \bar{\alpha}} \,D_{\bar{i}} \bar{M}_{\cal I} \,D_{\bar{\alpha}} \bar{L}^{\cal A} \, ,
\eea
imply both $D_i Z = 0$ and $D_{\alpha} Z = 0$ due to the special K\"{a}hler geometry of the complex and K\"{a}hler moduli spaces respectively. This gives us a strong motivation to investigate whether in (at least some cases of) $SU(3)\times SU(3)$ structure compactifications of type II strings the central charge $Z = e^{K/2} W$ for $N=1$ truncations can have the form (\ref{ZSU3sq}).

\subsection{Type II on $SU(3)\times SU(3)$ structure spaces} \label{SU3SU3}

Now we turn to type IIA/B string theory compactified on $SU(3)\times SU(3)$ structure spaces. Besides the technical motivation that we reached in the previous subsection, these compactifications are physically the most appropriate arena to explore the possibility for the existence of flux vacua attractors. The reason is that they give the most general type II flux compactifications with $N=1$ vacua \cite{GMPT}. In addition, it is natural to expect that such a generalization may be possible, since the deformation spaces of those manifolds have special K\"{a}hler structure \cite{NH,GLW1,GLW2,CB}.

\subsubsection{$SU(3)\times SU(3)$ structure compactifications}

Let us start by recalling some background material about $SU(3)\times SU(3)$ structure compactifications. For a more thorough review, see for example \cite{CB}; also, see \cite{KM}  for a detailed investigation showing that the susy conditions of the 4d effective action, in the presence of warp factors, indeed give solutions of the supersymmetry conditions of the full ten-dimensional theory.\footnote{It is interesting to note that, in addition, \cite{KM} studies the ten-dimensional interpretation of 4d non-perturbative effects, showing that the latter induce deformations of the generalized complex structure of the internal manifold. This new perspective also leads to an independent geometric derivation \cite{KM} of the superpotential for a mobile D3-brane, that was found earlier in \cite{mobile}.}

Manifolds with $SU(3)\times SU(3)$ structure are most naturally described in terms of generalized complex geometry \cite{NH, MG, FW}. The latter is a mathematical framework that provides a unifying description of various structures, existing on the tangent bundle of a manifold, by going to the sum of the tangent $T$ and cotangent $T^*$ bundles. For six-dimensional internal manifolds the natural structure group on $T\oplus T^*$ is $SO(6,6)$. Requiring that the manifold admit two globally defined spinors $\eta_1, \eta_2$ reduces the latter to $SU(3)\times SU(3)$ structure. This structure on $T\oplus T^*$ gives rise to several different structures on $T$. In particular, if $\eta_1$ and $\eta_2$ are proportional everywhere, then one obtains the more familiar $SU(3)$ structure. If $\eta_1$ and $\eta_2$ are never proportional, then one has an $SU(2)$ structure that is the intersection of the two $SU(3)$ structures defined by the two spinors. Clearly though, in general $\eta_1$ and $\eta_2$ need not be proportional (or orthogonal) everywhere; the angle between them can vary throughout the manifold. So the language of $SU(3)\times SU(3)$ structures is the most suitable one in order to encompass all possibilities.

An $SU(3)\times SU(3)$ structure is characterized by a pair of pure $SO(6,6)$ spinors $\Phi_+$ and $\Phi_-$, which can be viewed as elements of $\Lambda^{\bullet}T^*$ or, more precisely, as sums of even and odd forms respectively. They encode the geometric and B-field degrees of freedom of the internal manifold. Recall that, in the $SU(3)$ structure case, these are: $\Phi_+ = {\rm e}^{-(B+iJ)}$ and $\Phi_- = \Omega$\,, where $J$ and $\Omega$ are the defining 2- and 3-form. In general, however, $\Phi_-$ can contain 1-, 3- and 5-forms just like $\Phi_+$ contains 0-, 2-, 4- and 6-forms. In order to obtain an effective four-dimensional theory\footnote{At this point, that would be an $N=2$ theory since there are two internal spinors. We will discuss the $N=1$ vacua in the next subsubsection.}, one needs to expand the higher-dimensional fields (including $\Phi_{\pm}$) in terms of a finite basis of forms on the internal manifold and to keep only the light modes. However, for $SU(3)\times SU(3)$ structure manifolds (in fact, even for $SU(3)$ structure ones) the distinction between heavy and light modes is not straightforward, as the explicit construction of the appropriate basis of forms is not known in principle.\footnote{See \cite{AKP} and \cite{CKKLTZ}, though, for explicit examples in the cases of type IIA on nearly K\"{a}hler manifolds and on nilmanifolds and cosets, respectively.} Nevertheless, one can proceed by assuming the existence of a finite basis satisfying certain constraints, such that the resulting effective 4d theory is a consistent (gauged) $N=2$ supergravity \cite{GLW1,GLW2}. This approach has been quite fruitful so far and we adopt it in the following. Let us denote the set of odd basis forms as $\{ \alpha_{\cal I}, \beta^{\cal I} \}$ and the set of even basis forms as $\{ \omega_{\cal A}, \tilde{\omega}^{\cal A} \}$. As before, they are required to satisfy (\ref{symppair}); however, now the set $\{ \alpha_{\cal I}, \beta^{\cal I} \}$ contains 1-, 3- and 5-forms, unlike in Subsection \ref{HetnonKah} where it only contained 3-forms. 

Having introduced a finite basis of forms, we can decompose the pure spinors $\Phi_+$ and $\Phi_-$ as:
\be \label{PpPm}
\Phi_- = X^{\cal I} (z) \,\alpha_{\cal I} - G_{\cal I} (z) \,\beta^{\cal I} \,\, , \qquad
\Phi_+ = X^{\cal A} (t) \,\omega_{\cal A} - G_{\cal A} (t) \,\tilde{\omega}^{\cal A} \,\, .
\ee
As in the previous subsection, we have denoted by $z^i$ the coordinates of the moduli space ${\cal M}_-$ of deformations of $\Phi_-$ (for CY: the complex structure moduli) and by $t^{\alpha}$ the coordinates on the moduli space ${\cal M}_+$ of $\Phi_+$ deformations (for CY: the K\"{a}hler structure moduli). Clearly, the periods of $\Phi_{\pm}$ are defined by:
\be
X^{\cal I} = \int \la \Phi_- , \beta^{\cal I} \ra \, , \qquad G_{\cal I} = \int \la \Phi_- , \alpha_{\cal I} \ra 
\ee
and 
\be
X^{\cal A} = \int \la \Phi_+ , \tilde{\omega}^{\cal A} \ra \, , \qquad G_{\cal A} = \int \la \Phi_+ , \omega_{\cal A} \ra \, ,
\ee
where $\la \, ,\ra$ is the Mukai pairing (\ref{Mukai}). The moduli spaces ${\cal M}_-$ and ${\cal M}_+$ have special K\"{a}hler metrics with K\"{a}hler potentials \cite{GLW2}:
\be
K_- (z) = - \ln i \int \la \Phi_-, \bar{\Phi}_- \ra = - \ln i (\bar{X}^{\cal I} G_{\cal I} - X^{\cal I} \bar{G}_{\cal I})
\ee
and
\be
K_+ (t) = - \ln i \int \la \Phi_+, \bar{\Phi}_+ \ra = - \ln i (\bar{X}^{\cal A} G_{\cal A} - X^{\cal A} \bar{G}_{\cal A}) \,\, ,
\ee
respectively. As before, one can define covariantly holomorphic symplectic sections:
\bea \label{PhipPhimSec}
L^{\cal I} (z) = e^{\frac{K_-}{2}} X^{\cal I} &,& M_{\cal I} (z) = e^{\frac{K_-}{2}} G_{\cal I} \,\,\,\, , \nn \\
L^{\cal A} (t) = e^{\frac{K_+}{2}} X^{\cal A} &,& M_{\cal A} (t) = e^{\frac{K_+}{2}} G_{\cal A} \,\,\,\, ,
\eea
so that 
\be \label{LMnor}
i (\bar{L}^{\cal I} M_{\cal I} - L^{\cal I} \bar{M}_{\cal I}) = 1 \qquad {\rm and} \qquad i (\bar{L}^{\cal A} M_{\cal A} - L^{\cal A} \bar{M}_{\cal A}) = 1 \, .
\ee

Similarly to (\ref{HaFl}), the basis forms for $SU(3)\times SU(3)$ structure spaces are not closed in general. However, now the differential conditions they satisfy are \cite{GLW2}:\footnote{For later convenience, we adopt different sign conventions for $\tilde{e}_{\cal I}^{\cal A}$ and $\tilde{m}^{{\cal I A}}$ compared to \cite{GLW2}.} 
\bea \label{DiffCond}
{\cal D} \omega_{\cal A} &\sim& m_{\cal A}^{\cal I} \,\alpha_{\cal I} - e_{{\cal I} {\cal A}} \,\beta^{\cal I} \nn \\
{\cal D} \tilde{\omega}^{\cal A} &\sim& \tilde{m}^{{\cal I} {\cal A}} \,\alpha_{\cal I} - \tilde{e}^{\cal A}_{\cal I} \,\beta^{\cal I} \nn \\
{\cal D} \alpha_{\cal I} &\sim& - \tilde{e}_{\cal I}^{\cal A} \,\omega_{\cal A} + e_{{\cal I} {\cal A}} \,\tilde{\omega}^{\cal A} \nn \\
{\cal D} \beta^{\cal I} &\sim& - \tilde{m}^{{\cal I} {\cal A}} \,\omega_{\cal A} + m^{\cal I}_{\cal A} \,\tilde{\omega}^{\cal A} \,\, .
\eea
Here the symbol $\sim$ means equality only up to terms that vanish under the symplectic pairing (\ref{symppair}). The generalized 'derivative' operator ${\cal D}$ is defined by \cite{STW,CB}:
\be \label{GenDer}
{\cal D} = d - H \wedge \,- \,Q \cdot - \,R \llcorner \,\,\,\, ,
\ee
where $H$ is the NS 3-form flux and the operators $Q$ and $R$ act on a $p$-form $C$ as
\bea
(Q \cdot C)_{m_1...m_{p-1}} &=& Q^{n_1 n_2}{}_{[m_1} C_{|n_1n_2|m_2...m_{p-1}]} \,\,\, , \nn \\
(R \llcorner C)_{m_1...m_{p-3}} &=& R^{n_1n_2n_3}C_{n_1n_2n_3m_1...m_{p-3}} \,\,\, .
\eea
Hence ${\cal D}$ still maps even forms into odd forms and vice versa. The $Q$ and $R$ components in (\ref{GenDer}) appear when one considers non-geometric backgrounds. Finally, the constant charge matrices $e_{\cal I A}$, $m^{\cal I}_{\cal A}$, $\tilde{e}_{\cal I}^{\cal A}$ and $\tilde{m}^{\cal I A}$ in (\ref{DiffCond}) have to satisfy the constraints:
\bea
&&\hspace*{-1.5cm}\tilde{m}^{\cal I A} m^{\cal J}_{\cal A} - m^{\cal I}_{\cal A} \tilde{m}^{\cal J A} = 0 \, , \qquad \hspace*{0.03cm}\tilde{e}_{\cal I}^{\cal A} e_{\cal J A} - e_{\cal I A} \tilde{e}^{\cal A}_{\cal J} = 0 \, , \quad \hspace*{0.26cm}\tilde{e}_{\cal I}^{\cal A} m^{\cal J}_{\cal A} - e_{\cal I A} \tilde{m}^{\cal J A} = 0 \nn \\
&&\hspace*{-1cm}\tilde{m}^{\cal I A} \tilde{e}_{\cal I}^{\cal B} - \tilde{e}^{\cal A}_{\cal I} \tilde{m}^{\cal I B} = 0 \, , \qquad \hspace*{-0.15cm}m^{\cal I}_{\cal A} e_{\cal I B} - e_{\cal I A} m^{\cal I}_{\cal B} = 0 \, , \qquad m^{\cal I}_{\cal A} \tilde{e}^{\cal B}_{\cal I} - e_{\cal I A} \tilde{m}^{\cal I B} = 0 \, ,
\eea
in order for the nilpotency condition ${\cal D}^2 = 0$ to hold. Let us also note that \cite{GLW2} argued that, in order to turn on all charges in the total charge matrix (\ref{Qtot}), one has to consider non-geometric compactifications \cite{CH, STW}. The latter differ from the geometric ones in that their transition functions contain string dualities, like T-duality. Nevertheless, the naive supergravity treatment still gives the correct low-energy effective theory \cite{GLW2}; see also \cite{MPT}. For more details on the relation between $SU(3)\times SU(3)$ structure geometric and non-geometric compactifications, see \cite{nonGeom}.\footnote{A beautiful thorough investigation of this issue was performed in \cite{GMPW}, which appeared after the first version of this paper. More precisely, this work showed that those of the charges $e_{\cal I A}$, $\tilde{e}_{\cal I}^{\cal A}$, $m^{\cal I}_{\cal A}$ and $\tilde{m}^{\cal I A}$ on the right hand side of (\ref{DiffCond}), which are due to non-zero $Q$ and $R$ in (\ref{GenDer}), can always be gauged away locally. (I.e., locally one can always choose a basis, such that they vanish.) However, in contrast to the geometric case, for non-geometric backgrounds there are obstructions for such a choice globally. For more details, see \cite{GMPW}.}

As already mentioned above, generically type II on $SU(3)\times SU(3)$ structure spaces leads to an $N=2$ effective theory as there are two internal spinors. In the following, we will be interested in truncations that preserve only $N=1$ susy in four dimensions.  

\subsubsection{$N=1$ truncation}

An obvious way of obtaining an $N=1$ truncation is to consider type II orientifold compactifications on $SU(3)\times SU(3)$ structure spaces. The resulting 4d effective theory was studied in \cite{BG}. However, the $N=2 \rightarrow N=1$ truncation does not have to come from orientifolding, as pointed out in \cite{CB}. It could be the result of a spontaneous partial  supersymmetry breaking. I.e., in such cases the $N=1$ theories provide low-energy effective descriptions around $N=1$ vacua that break $N=2$ spontaneously. Since the concrete mechanism, leading to the $N=1$ truncation, is irrelevant for our purposes, we will not dwell on that any further.

Let the $N=1$ susy parameter $\varepsilon$ be given by the linear combination $\varepsilon = a \varepsilon_1 + \bar{b} \varepsilon_2$ of the two $N=2$ parameters $\varepsilon_{1}$ and $\varepsilon_2$, where $a$ and $b$ are complex constants such that $|a|^2+|b|^2 = 1$. Also, let us first concentrate on type IIA. Then \cite{CB} finds that:
\be \label{KW}
e^{K/2} W = \frac{i}{4 \bar{a} b} e^{\frac{K_+}{2} + 2\varphi} \int \la \Phi_+ , {\cal D} \Pi_- + \frac{1}{\sqrt{2}} G^{fl} \ra \,\, ,
\ee
where 
\be
\Pi_- = \frac{1}{\sqrt{2}} A^{odd} + i {\rm Im} (C \Phi_-) \qquad {\rm with} \qquad C = \sqrt{2} ab e^{-\phi} \,\, .
\ee
Here $\phi$ is the ten-dimensional dilaton, $\varphi$ is the 4d one given by $e^{-2\varphi} = \int e^{-2\phi} vol_6$ and, finally, the RR potential $A^{odd}$ and RR flux $G^{fl}$ are defined by $G=G^{fl} + {\cal D} A^{odd}$, where $G=G_0+G_2+G_4+G_6$ is the sum of all internal RR field strengths.\footnote{Note that $G$ is related to the sum $F$ of the usual modified field strengths, that enter the ten-dimensional supergravity action, via $F = e^B G$.} Using that \cite{BG,DGKT}:
\be \label{Khat}
\hat{K}_- = -2 \ln i \int \la \Pi_- , \bar{\Pi}_- \ra = 4 \varphi \,\, ,
\ee
where $\hat{K}_-$ is the K\"{a}hler potential for the space of complex scalars that arise in the $\Pi_-$ expansion\footnote{Strictly speaking, (\ref{Khat}) was rigorously derived only for $SU(3)$ structure compactifications \cite{BG}. However, it is natural to expect that it holds for $SU(3)\times SU(3)$ structure compactifications as well, since results for the $SU(3)$ structure case, when expressed in terms of the pair $(\Phi_+, \Phi_-)$ instead of just $(J, \Omega)$, usually extend to the full $SU(3)\times SU(3)$ structure case; see \cite{GLW1} and \cite{GLW2}, for example.}, we recognize the total $N=1$ K\"{a}hler potential in (\ref{KW}) to be $K = K_+ + \hat{K}_-$. 

For simplicity, let us consider the case of vanishing RR fluxes. Then the $N=1$ superpotential is:
\be
W = c \int \la \Phi_+ , {\cal D} \Pi_- \ra \,\, ,
\ee 
where we have denoted $c = \frac{i}{4\bar{a}b}$. Expanding $\Pi_-$ on the basis of odd forms,
\be \label{PimExpan}
\Pi_- = \hat{X}^{\cal I} \alpha_{\cal I} - \hat{G}_{\cal I} \beta^{\cal I} \, ,
\ee
and using (\ref{DiffCond}), (\ref{symppair}) and the $\Phi_+$ expansion in (\ref{PpPm}), one can write the superpotential as:
\be
W = c (\hat{X}^{\cal I} e_{\cal I A} X^{\cal A} - \hat{G}_{\cal I} m^{\cal I}_{\cal A} X^{\cal A} - \hat{X}^{\cal I} \tilde{e}_{\cal I}^{\cal A} G_{\cal A} + \hat{G}_{\cal I} \tilde{m}^{\cal I A} G_{\cal A}) \,\, .
\ee
Introducing
\be \label{LhMh}
\hat{L}^{\cal I} = e^{\frac{\hat{K}_-}{2}} \hat{X}^{\cal I} \qquad {\rm and} \qquad \hat{M}_{\cal I} = e^{\frac{\hat{K}_-}{2}} \hat{G}_{\cal I} \,\, ,
\ee
we find that the central charge acquires the form:
\be \label{ZSU3SU3}
Z = e^{K/2} W = c \left[e_{{\cal I} {\cal A}} \hat{L}^{\cal I} L^{\cal A} - m_{\cal A}^{\cal I} \hat{M}_{\cal I} L^{\cal A} - \tilde{e}_{\cal I}^{\cal A} \hat{L}^{\cal I} M_{\cal A} + \tilde{m}^{{\cal I} \cal A} \hat{M}_{\cal I} M_{\cal A}\right] \, .
\ee

The last expression looks exactly like (\ref{ZSU3sq}). Unfortunately though, there is a key difference. Namely, the metric determined by the K\"{a}hler potential (\ref{Khat}) is {\it not} special K\"{a}hler. In particular,
\be \label{Kmhat}
\hat{K}_- = - 2 \,\ln i \left(\bar{\hat{X}}^{\cal I} \hat{G}_{\cal I} - \hat{X}^{\cal I} \bar{\hat{G}}_{\cal I}\right)
\ee 
implies that 
\be \label{hatNorm}
i \left( \bar{\hat{L}}^{\cal I} \hat{M}_{\cal I} - \hat{L}^{\cal I} \bar{\hat{M}}_{\cal I} \right) = e^{\frac{\hat{K}_-}{2}} \,\, ,
\ee
unlike the normalizations (\ref{LMnor}). In addition, the coordinates $\hat{X}^{\cal I}$, $\hat{G}_{\cal I}$ are not projective and are actually independent \cite{BG, DGKT}. In other words, the set of supersymmetry conditions is:
\be \label{susyPim}
D_{\alpha} Z = 0 \, , \qquad D_{\hat{X}^{\cal I}} Z = 0 \, , \qquad D_{\hat{G}_{\cal I}} Z = 0 \, .
\ee

Whereas, according to our previous considerations, the first condition is satisfied by the conjecture (\ref{GenKahlConj}) with indices $i$ running over $\{ \hat{X}^{\cal I}, \hat{G}_{\cal I}\}$ together with the substitutions $\Omega \rightarrow \Pi_-$ and $e^{-J_c} \rightarrow \Phi_+$ (plus taking into account the proper normalization for the $\Pi_-$ sections, as we will see below), the last two conditions in (\ref{susyPim}) are not. Nevertheless, the form of $\hat{K}_-$ is very specific. This, together with the fact that $K_+$ does determine a special K\"{a}hler metric, makes it worth investigating whether there are any conditions under which $D_{\hat{X}^{\cal I}} Z = 0$ and $D_{\hat{G}_{\cal I}} Z = 0$ can still be satisfied as a result of the relevant attractor equations. We will see now that the answer to this question is positive. More precisely, we will show that {\it all} of the susy conditions in (\ref{susyPim}) are implied by the appropriate attractor equations when one considers {\it only} Minkowski vacua.

\subsubsection{Minkowski attractors}

Let us introduce the analogue of (\ref{DoubleSec}) for the present case:
\be
{\cal U} = e^{(\hat{K}_- + K_+)/2} (\,\Pi_- \otimes \Phi_+) \,\, .
\ee
Using (\ref{PimExpan}), (\ref{LhMh}) and (\ref{hatNorm}), one can easily compute that ${\cal U}$ is normalized as:
\be
\int \la \bar{{\cal U}} , {\cal U} \ra = (\bar{\hat{L}}^{\cal I} \hat{M}_{\cal I} - \hat{L}^{\cal I} \bar{\hat{M}}_{\cal I}) (\bar{L}^{\cal A} M_{\cal A} - L^{\cal A} \bar{M}_{\cal A}) = - e^{\frac{\hat{K}_-}{2}} \,\, .
\ee
It is also easy to see that (\ref{ZSU3SU3}) can be written as:
\be
Z = c \int \la \hat{{\cal Q}} , {\cal U} \ra \,\, ,
\ee
where $\hat{{\cal Q}}$ is given in (\ref{Qgen}) with the meaning of the indices ${\cal I}$ being as in the current Subsection \ref{SU3SU3}. Hence the analogue of (\ref{GenKahlConj}) is:
\be \label{SU3SU3Attr}
\hat{{\cal Q}} = - \frac{2}{c} \,e^{-\frac{\hat{K}_-}{2}} \,{\rm Re} ( \bar{Z} {\cal U} + g^{\hat{i} \bar{\hat{j}}} g^{\alpha \bar{\beta}} \,D_{\hat{i}} D_{\alpha} {\cal U} \,\bar{D}_{\bar{\hat{j}}} \bar{D}_{\bar{\beta}} \bar{Z} ) \,\, ,
\ee
where $\hat{i}$, $\hat{j}$ run over the set of independent variables $\{\hat{X}^{\cal I}, \hat{G}_{\cal I} \}$ and, as usual, the right-hand side of (\ref{SU3SU3Attr}) is understood to be evaluated at the susy minima.

This proposal implies the susy condition $D_{\beta} Z = 0$ similarly to the considerations at the end of Subsection \ref{HetnonKah}. Indeed:
\bea
D_{\beta} Z &=& c\left[ e_{\cal I A} \hat{L}^{\cal I} \,D_{\beta} L^{\cal A} - m^{\cal I}_{\cal A} \hat{M}_{\cal I} \,D_{\beta} L^{\cal A} - \tilde{e}^{\cal A}_{\cal I} \hat{L}^{\cal I} \,D_{\beta} M_{\cal A} + \tilde{m}^{\cal I A} \hat{M}_{\cal I} \,D_{\beta} M_{\cal A} \right] \nn \\
&=& - e^{-\frac{\hat{K}_-}{2}} \left[ Z (\hat{L}^{\cal I} \bar{\hat{M}}_{\cal I} - \bar{\hat{L}}^{\cal I} \hat{M}_{\cal I}) (\bar{M}_{\cal A} \,D_{\beta} L^{\cal A} - \bar{L}^{\cal A} \,D_{\beta} M_{\cal A}) \right. \nn \\
&+& \left. Z^{\bar{i} \bar{\alpha}} (\hat{L}^{\cal I} \,D_{\bar{i}} \bar{\hat{M}}_{\cal I} - \hat{M}_{\cal I} \,D_{\bar{i}} \bar{\hat{L}}^{\cal I}) (D_{\beta} L^{\cal A} \,D_{\bar{\alpha}} \bar{M}_{\cal A} - D_{\bar{\alpha}} \bar{L}^{\cal A} \,D_{\beta} M_{\cal A}) \right] = \,0 \,\, ,
\eea
where both terms vanish due to the special geometry properties of $(L^{\cal A}, M_{\cal A})$ alone. Let us now concentrate on the remaining two susy conditions: $\left( \pd_{\hat{X}^{\cal I}} + (\pd_{\hat{X}^{\cal I}} \hat{K}_-) \right) W = 0$\,, or equivalently $D_{\hat{X}^{\cal I}} Z \equiv \left( \pd_{\hat{X}^{\cal I}}+\frac{1}{2} (\pd_{\hat{X}^{\cal I}} \hat{K}_-) \right) Z = 0$\,, and similarly for $\hat{G}_{\cal I}$.\footnote{Of course, the equivalence between $\left( \pd_{\hat{X}^{\cal I}} + (\pd_{\hat{X}^{\cal I}} \hat{K}_-) \right) W = 0$ \,and $\left( \pd_{\hat{X}^{\cal I}}+\frac{1}{2} (\pd_{\hat{X}^{\cal I}} \hat{K}_-) \right) Z = 0$ is valid for $e^{(K_+ + \hat{K}_-)/2}$ nonvanishing or, in other words, for a total K\"{a}hler potential $K= K_+ + \hat{K}_-$ that does not go to $-\infty$ anywhere on moduli space.} From the expression for the central charge in (\ref{ZSU3SU3}), we find that:
\be \label{DhatXZ}
D_{\hat{X}^{\cal I}} Z = c \,( e_{\cal I A} L^{\cal A} - \tilde{e}_{\cal I}^{\cal A} M_{\cal A} ) \,e^{\frac{\hat{K}_-}{2}} + ( \pd_{\hat{X}^{\cal I}} \hat{K}_- ) Z \,\, .
\ee
Now, (\ref{Kmhat}) implies that
\be \label{Kmder}
\pd_{\hat{X}^{\cal I}} \hat{K}_- = \frac{2 \,\bar{\hat{G}}_{\cal I}}{i (\bar{\hat{X}}^{\cal J} \hat{G}_{\cal J} - \hat{X}^{\cal J} \bar{\hat{G}}_{\cal J})} = 2 \bar{\hat{M}}_{\cal I} \,\, ,
\ee
whereas the expressions for $e_{\cal I A}$ and $\tilde{e}_{\cal I}^{\cal A}$ in (\ref{SU3SU3Attr}) lead to:
\be \label{eet}
e_{\cal I A} L^{\cal A} - \tilde{e}_{\cal I}^{\cal A} M_{\cal A} = - \frac{i}{c} \,e^{-\frac{\hat{K}_-}{2}} Z \bar{\hat{M}}_{\cal I} \,\, .
\ee
Note that in the last result all $DDZ$ terms canceled due to the special K\"{a}hler geometry of the $\Phi_+$ moduli space, more precisely due to $L^{\cal A} D_{\alpha} M_{\cal A} - M_{\cal A} D_{\alpha} L^{\cal A} = 0$ (see (\ref{SpRel})) together with the second condition in (\ref{LMnor}). Finally, substituting (\ref{Kmder}) and (\ref{eet}) in (\ref{DhatXZ}), we obtain:
\be
D_{\hat{X}^{\cal I}} Z = (2 - i) Z \bar{\hat{M}}_{\cal I} \,\, .
\ee
Similarly, one also finds that:
\be
D_{\hat{G}_{\cal I}} Z = c \,(-m_{\cal A}^{\cal I} L^{\cal A} + \tilde{m}^{\cal I A} M_{\cal A}) \,e^{\frac{\hat{K}_-}{2}} + (\pd_{\hat{G}_{\cal I}} \hat{K}_-) Z = (i - 2) Z \bar{\hat{L}}^{\cal I} \,\, .
\ee
Clearly then, in general both the $\hat{X}^{\cal I}$ and the $\hat{G}_{\cal I}$ supersymmetry  conditions are not satisfied. However, for Minkowski vacua $Z=0$ and so in such a case one has $D_{\hat{X}^{\cal I}} Z = 0$ and $D_{\hat{G}_{\cal I}} Z = 0$. 

For type IIB on $SU(3)\times SU(3)$ structure spaces the situation is the same as above but with the exchange of indices ${\cal I} \leftrightarrow {\cal A}$\,, since the IIB case is obtained from the IIA one by the substitutions $\Phi_+ \rightarrow \Phi_-$\,, \,$\Pi_- \rightarrow \Pi_+ = \frac{1}{\sqrt{2}} A^{ev} + i \,{\rm Im} (C \Phi_+)$ and $G_0+G_2+G_4+G_6 \,\rightarrow \,G_1+G_3+G_5$ in (\ref{KW}) \cite{GLW1,GLW2,BG}. More precisely, the IIB attractors for supersymmetric Minkowski vacua (with vanishing RR fluxes) are:
\be \label{IIBSU3SU3}
\hat{{\cal Q}} = - \frac{2}{c} \,e^{-\frac{\hat{K}_+}{2}} \,{\rm Re} ( g^{i \bar{j}} g^{\hat{\alpha} \bar{\hat{\beta}}} \,D_i D_{\hat{\alpha}} \tilde{{\cal U}} \,\bar{D}_{\bar{j}} \bar{D}_{\bar{\hat{\beta}}} \bar{Z} ) \,\, ,
\ee
where
\be
\tilde{{\cal U}} = e^{(K_- + \hat{K}_+)/2} (\,\Phi_-\otimes \Pi_+)
\ee
with
\be
\hat{K}_+ = -2 \,\ln i \int \la \,\Pi_+ \,, \,\bar{\Pi}_+ \ra = - 2 \,\ln i \left( \bar{\hat{X}}^{\cal A} \hat{G}_{\cal A} - \hat{X}^{\cal A} \bar{\hat{G}}_{\cal A} \right)
\ee
and indices $\hat{\alpha}$, $\hat{\beta}$ running over the independent variables $\hat{X}^{\cal A}$ and $\hat{G}_{\cal A}$.

In conclusion, we have found that there are attractor equations for $N=1$ Minkowski vacua of type II compactified on $SU(3)\times SU(3)$ structure spaces.

\section{Conclusions and discussion}

We explored in detail the possibility for the existence of attractor equations in flux compactifications with $N=1$ vacua. We filled a gap in the existing literature by verifying analytically that the flux vacua attractors of \cite{RK}, for the case of type IIB on CY(3) orientifolds, automatically lead to solutions of all relevant supersymmetry conditions.\footnote{Of course, we mean the conditions that determine supersymmetric minima only with respect to the complex structure moduli and axion-dilaton, since the flux superpotential in this case does not depend on the K\"{a}hler moduli.} Although this is just a consistency check regarding the derivation of \cite{RK}, it is necessary in order to understand how to generalize the flux vacua attractors beyond CY compactifications. 

We investigated possible generalizations for the heterotic string on $SU(3)$ structure and for type IIA/B on $SU(3)\times SU(3)$ structure and in both cases found flux vacua attractors for $N=1$ {\it Minkowski} vacua only.\footnote{Clearly, here the statement that the vacua are Minkowski is a restriction only for type IIA/B since, as we have already mentioned, the heterotic string does not have susy AdS vacua at the classical level.} Along the way, we also showed that a previous attractor proposal, for the case of heterotic non-K\"{a}hler compactifications with one K\"{a}hler modulus, actually fails, whereas another proposal has to be slightly modified in order to give heterotic attractors for arbitrary number of K\"{a}hler moduli. Our method, however, does not address the question whether {\it all} supersymmetric Minkowski vacua can be obtained from our attractor equations. Clearly, answering this question is an important component of the reformulation of the minimization of the relevant scalar potential into the problem of solving an appropriate system of attractor equations. Even more important is to understand whether there is a more conceptual explanation, as opposed to the technical one provided in this paper, of why these attractors do not give AdS vacua. This could, perhaps, be related to the need for a better understanding, in the context of generalized compactifications, of the cohomology decomposition of the relevant generalized flux, in the vein of \cite{CDF,DD,MS}. 

We should also comment on an apparent degeneracy of the Minkowski vacua, that at first sight might seem to be implied by our attractor equations. Namely, since the attractors (\ref{GenKahlConj}), (\ref{SU3SU3Attr}) and (\ref{IIBSU3SU3}) are only valid for Minkowski vacua (i.e, for $Z=0$), it may seem that one could remove the normalization factor $e^{-\frac{\hat{K}_-}{2}}$ in (\ref{SU3SU3Attr}), as this does not spoil the susy conditions; the same goes for the factor of $e^{-\frac{\hat{K}_+}{2}}$ in (\ref{IIBSU3SU3}). However, we believe that this freedom is spurious, since both the $Z$ and the $DDZ$ terms should have a common origin in the above mentioned cohomology decomposition at an arbitrary point in moduli space (not only at the supersymmetric extrema). Hence, the factors, needed for proper normalization at a generic $Z\neq 0$ point of moduli space, should also be present at points, in which $Z=0$.

Let us also point out, that the Minkowski attractors for type II on $SU(3)\times SU(3)$ strucure are not a conceptually trivial (due to string dualities) consequence of the Minkowski attractors for the heterotic string on $SU(3)$ structure. Part of the reason is that not much is known about the existence of a heterotic-type II duality in the presence of fluxes and/or non-K\"{a}hler compactifications. In fact, the only evidence we are aware of is in the case of heterotic on $K3\times T^2$ with flux, the dual being type IIA (M-theory) on $SU(3)$ structure \cite{HIID}. More importantly, however, even in the well-understood case of CY compactifications without flux, dualities require a particular fibration structure of the internal manifolds\footnote{For example, the familiar heterotic - F-theory duality operates only for the heterotic on an elliptically fibered CY(3) and F-theory on an elliptically fibered CY(4), whose base is a $\bb{P}^1$ fibration over the base of the heterotic side CY(3).}, whereas our considerations did not. It is also worth noting that physically interesting compactfications of heterotic or type II strings may be exactly such that they do not have a dual, as recently exemplified by the work of \cite{Vafa}, in which the F-theory compactifications of phenomenological interest are precisely those {\it without} heterotic duals. In addition, from our studies it is also clear that there is a difference at the technical level between the Minkowski attractors for the heterotic and for the type IIA/B strings. Namely, in the heterotic case the attractor equations (\ref{GenKahlConj}) themselves led to the vanishing of the central charge, as we saw in Subsection \ref{HetMinkAttr}. On the other hand, in the type II case we had to put $Z=0$ by hand in (\ref{SU3SU3Attr}) in order to satisfy the susy conditions. Clearly, it would be desirable to understand the reason behind this difference at a more conceptual level. The key should be the fact that type II, unlike the heterotic string, has classical supersymmetric AdS flux vacua and those should, probably, be encoded by attractor equations that arise from a more general cohomological decomposition of the generalized flux than the one we considered here. In particular, the attractor equations could contain terms proportional to $D_{\hat{i}} D_{\hat{j}} Z$, for example, in addition to those in (\ref{SU3SU3Attr}), and respectively $D_{\hat{\alpha}} D_{\hat{\beta}}Z$ for the case of (\ref{IIBSU3SU3}), since the relevant K\"{a}hler potentials, $\hat{K}_{\pm}$, are {\it not} special K\"{a}hler. We hope to come back to this in the near future.

Other open issues include the following. An immediate open problem is to investigate whether one can extend the attractors, that we found for type II on $SU(3)\times SU(3)$ structure, to the case of non-vanishing RR fluxes. In particular, it would be very interesting to see whether turning on RR fluxes can lead to supersymmetric {\it AdS} attractor equations. It would also be of great interest to address the possibility for existence of nonsupersymmetric attractors, both for the heterotic on $SU(3)$ structure and for type II on $SU(3)\times SU(3)$ structure, as that would provide another tool for studying de Sitter vacua. It is also worth pointing out that there might be flux vacua attractors in a broader context than the kind of compactifications that we have studied here. The inspiration for this suggestion comes from recent studies of BH attractors in six and seven dimensions \cite{Saidi}. It is interesting to explore whether the latter have analogy in the realm of flux vacua attractors. Finally, it is, of course, of great importance to investigate how much one can learn about moduli stabilization from the already found Minkowski flux vacua attractors. This is especially interesting, given the difficulty in finding type II Minkowski vacua with stabilized moduli; for recent progress in that direction using different methods, see \cite{BBVW, MPT}.

\section*{Acknowledgements}

I would like to thank B. de Wit, K. Goldstein, A. Marrani and especially S. Vandoren for useful discussions. I am very grateful to A. Ceresole for reading the draft and providing valuable feedback and to P.A. Grassi for positive interaction. I also thank G. Dall'Agata for correspondence. In addition, I am grateful to the Univ. of Cincinnati and to the 6th Simons workshop in Mathematics and Physics, Stony Brook 2008, for hospitality during the completion of this work. My research is supported in part by the EU RTN network MRTN-CT-2004-005104 and INTAS contract 03-51-6346.

\appendix

\section{Appendix} \label{app}
\setcounter{equation}{0}

To understand why not every solution of $D Z = 0$ is compatible with (\ref{Znew}), let us first look at the simpler case of the BH attractors (\ref{AttrEqBH}).

As we saw in Subsection \ref{ProofGen}, substituting
\be \label{OldAttr}
p^{\Lambda} = i (\bar{Z} L^{\Lambda} - Z \bar{L}^{\Lambda}) \qquad {\rm and} \qquad q_{\Lambda} = i (\bar{Z} M_{\Lambda} - Z \bar{M}_{\Lambda})
\ee
into
\be \label{DZAp}
D_i Z = q_{\Lambda} D_i L^{\Lambda} - p^{\Lambda} D_i M_{\Lambda} \, ,
\ee
one finds that $D_i Z = 0$ due to special K\"{a}hler geometry. However, one can easily check that the following expressions for the charges:
\be \label{AltSol}
p^{\Lambda} = \bar{Z} L^{\Lambda} + Z \bar{L}^{\Lambda} \qquad {\rm and} \qquad q_{\Lambda} = \bar{Z} M_{\Lambda} + Z \bar{M}_{\Lambda}
\ee 
also imply $D_i Z = 0$. Indeed, substituting (\ref{AltSol}) in (\ref{DZAp}), we find:
\be
D_i Z = \bar{Z} (M_{\Lambda} D_i L^{\Lambda} - L^{\Lambda} D_i M_{\Lambda}) = \bar{Z} L^{\Lambda} ({\cal N}_{\Lambda \Sigma} - \bar{{\cal N}}_{\Lambda \Sigma}) D_i L^{\Sigma} = 0 \, ,
\ee
where we have used (\ref{MNL}) and (\ref{SpRel}). Despite this, (\ref{AltSol}) are not alternative attractor equations as they are not compatible with
\be \label{ZAp}
Z = q_{\Lambda} L^{\Lambda} - p^{\Lambda} M_{\Lambda} \, .
\ee
Namely, substituting (\ref{AltSol}) into the right-hand side of (\ref{ZAp}) and using (\ref{Norm}), one obtains $iZ$ instead of $Z$.\footnote{Another example of fake solutions is the following: Obviously, one can generate an infinite number of solutions of $D_i Z = 0$ by multiplying the right-hand side of (\ref{OldAttr}) by an arbitrary constant (or even a function) $c$. However, those putative solutions would not be compatible with (\ref{ZAp}), unless $c=1$.} 

Clearly, the reason there can be spurious solutions of the susy conditions is that both $ q_{\Lambda} D_i L^{\Lambda} - p^{\Lambda} D_i M_{\Lambda} = 0$ and $Z = q_{\Lambda} L^{\Lambda} - p^{\Lambda} M_{\Lambda}$ are linear algebraic equations in the charges and in that sense can be viewed as independent conditions. The lesson from this discussion is that, in principle, one has to verify not only that the new attractor equations (\ref{AE2}) imply automatically $D Z = 0$ but also that they are compatible with the expression for $Z$ in (\ref{Znew}). As we already mentioned in the main text, this is not at all unexpected. Despite that, it is worth providing the explicit check here; this exercise will be rather useful for the considerations of Subsection \ref{SingleKahMod}.

Consider the right-hand side of (\ref{Znew}) with (\ref{AE2}) substituted:
\bea \label{ApComp}
(q_{fa} - \tau q_{ha}) L^a - (p_f^a \hspace*{-0.2cm}&-& \hspace*{-0.2cm}\tau p_h^a) M_a \hspace*{0.2cm}= \hspace*{0.2cm}(\tau \bar{Z} M_a + \bar{\tau} Z \bar{M}_a) L^a - \tau (\bar{Z} M_a + Z \bar{M}_a) L^a \\
&-& (\tau \bar{Z} L^a +\bar{\tau} Z \bar{L}^a) M_a + \tau (\bar{Z} L^a + Z \bar{L}^a) M_a \nn \\
&+& (\bar{\tau} \bar{Z}^{\underline{0} I} D_I M_a + \tau Z^{\underline{0} I} \bar{D}_I \bar{M}_a) L^a - \tau (\bar{Z}^{\underline{0} I} D_I M_a + Z^{\underline{0} I} \bar{D}_I \bar{M}_a) L^a \nn \\
&-& (\bar{\tau} \bar{Z}^{\underline{0} I} D_I L^a + \tau Z^{\underline{0} I} \bar{D}_I \bar{L}^a) M_a + \tau (\bar{Z}^{\underline{0} I} D_I L^a + Z^{\underline{0} I} \bar{D}_I \bar{L}^a) M_a \nn \\
&=& Z (\tau - \bar{\tau}) (\bar{L}^a M_a - L^a \bar{M}_a) - (\tau - \bar{\tau}) \bar{Z}^{\underline{0}I} (L^a D_I M_a - M_a D_I L^a) \nn \\
&=& -i Z (\tau - \bar{\tau}) \,e^{K_{(\tau)}} - (\tau - \bar{\tau}) \bar{Z}^{\underline{0} i} L^a (\bar{{\cal N}}_{ab} - {\cal N}_{ab}) D_i L^b \,\, = \,\, Z \,\,\,\, , \nn
\eea
where we have used that $L^a (\bar{{\cal N}}_{ab} - {\cal N}_{ab}) D_i L^b = 0$, as shown in Section \ref{AttrProof} (see (\ref{SpRel})), and also that $i(\bar{L}^a M_a - L^a \bar{M}_a) = e^{K_{(\tau)}}$ with $K_{(\tau)} = - \ln [-i (\tau - \bar{\tau})]$. This completes the proof.

\end{document}